\documentclass[fleqn,usenatbib]{mnras}

\usepackage{newtxtext,newtxmath}
\usepackage{ulem}

\usepackage[T1]{fontenc}
\usepackage{tablefootnote}
\usepackage{booktabs}
\usepackage{threeparttable}
\usepackage{threeparttablex}
\usepackage{thmtools}
\usepackage{orcidlink}
\DeclareRobustCommand{\VAN}[3]{#2}
\let\VANthebibliography\thebibliography
\def\thebibliography{\DeclareRobustCommand{\VAN}[3]{##3}\VANthebibliography}

\usepackage{graphicx}	
\usepackage{amsmath}	

\title[MeerKAT chromospheric stars]{Chromospherically Active Stars in the MeerKAT Galactic Plane Survey}

\author[O. D. Egbo et al.]{
Okwudili D. Egbo\orcidlink{0009-0001-0232-3968}$^{1,2,}$\thanks{E-mail: egbo@saao.ac.za (ODE)},
David A. H. Buckley\orcidlink{0000-0002-7004-9956}$^{1,2}$,
Paul J. Groot\orcidlink{0000-0002-4488-726X}$^{1,2,3}$,
Orapeleng Mogawana\orcidlink{0000-0002-2261-0382}$^{1,2}$,
\newauthor 
Anke van Dyk\orcidlink{0000-0003-0397-4655}$^{1,2}$,
Yusuke Tampo\orcidlink{0000-0003-0196-3936}$^{1,2}$,
Simon de Wet\orcidlink{0000-0003-2449-1329}$^{4}$
\\
$^{1}$Department of Astronomy, University of Cape Town, Private Bag X3, Rondebosch 7701, South Africa\\
$^{2}$South African Astronomical Observatory, PO Box 9, Observatory 7935, South Africa\\
$^{3}$Department of Astrophysics/IMAPP, Radboud University, PO 9010, NL-6500 GL Nijmegen, the Netherlands\\
$^{4}$DTU Space, Technical University of Denmark, Building 327, Elektrovej, 2800 Kgs. Lyngby, Denmark
}

\date{Accepted XXX. Received YYY; in original form ZZZ}

\pubyear{\the\year{}}

\begin{document}
\label{firstpage}
\pagerange{\pageref{firstpage}--\pageref{lastpage}}
\maketitle

\begin{abstract}
Stellar magnetic activity has long been established to occur in subsets of late-type main-sequence stars and manifests itself in the form of chromospheric and coronal activity. 
This activity is driven by magnetic fields generated by dynamo mechanisms associated with stellar rotation and convection, which channel and dissipate magnetic energy to the outer atmospheric layers. 
Optical spectroscopic follow-up was carried out for 99 radio-X-ray coronally active stars of G-, K-, and M-spectral types in the SARAO MeerKAT Galactic plane survey point sources. There are 46 stars that exhibited chromospheric activity.
Specifically, 39 stars show Ca {\sc ii} H\&K emission lines, whereas 18 show Ca {\sc ii} IRT, 11 show both H\&K and IRT lines. Of the 39 stars displaying H\&K emission, 37 show activity levels greater than the Solar minimum activity. Their X-ray to bolometric luminosity ratios are within the coronal thresholds of $> -4.5$. In addition, we explored optical photometry to investigate the rotation information and found that these stars show short rotation periods; in comparison with activity levels, the activity levels dampen as the rotation periods increase. We also analysed optical light curves for  a subset of stars, and they show flaring behaviour at the same epoch as enhanced chromospheric activity, indicating the flares and chromospheric emission are manifestations of the same underlying magnetic activity process. 
\end{abstract}




\section{Introduction}\label{sec:intro}

    Stellar atmospheres comprise the photosphere, the chromosphere and the corona, with the chromosphere sandwiched between the photosphere and the corona \citep{deGrijs:2021}. As stars evolve, heightened chromospheric activity is linked to the strength of the magnetic field in the outer layers of a star. The magnetic field strength depends on stellar mass and rotation rate, with faster rates associated with either a young age or the presence of a binary companion \citep{Hall:2008}.

    Late-type stars are known to exhibit active chromospheres \citep{Fekel:1986}, which are driven by rotation enhanced magnetic activity.
    The Ca {\sc ii} H \& K lines at 3968.5 \AA\, (H) and 3933.7 \AA\, (K), are the primary diagnostics of stellar chromospheric activity \citep{Benz:2010}. Their use can be traced back to early stellar spectroscopy in the mid-20th century, when Wilson initiated the monitoring of chromospheric activity in late-type stars with the 2.5 m telescope at the Mount Wilson Observatory to monitor the Ca {\sc ii} H \& K emission lines \citep{Wilson:1968, Vaughan:1978, Vaughan:1980, Sowmya:2023}. Additionally, Vaughan used another 1.5 m telescope at the same Observatory for the continuous observing and tracing of the lines in active stellar systems \citep{Vaughan:1978, Vaughan:1980, Duncan:1991}. Both surveys yielded significant results, such that the Mount Wilson, $S$-index was adopted as a standard quantity for indicating the strength of chromospheric activities in stars \citep{Duncan:1991}. Henceforth, these lines became a fundamental linkage between optical and non-thermal phenomena in stars, enabling studies of stellar activity cycles analogous to the solar 11-year cycle \citep{Wilson:1978}.

    The connection between Ca {\sc ii} H \& K emission and stellar magnetic activity is now well understood within the framework of dynamo-generated magnetic fields. The line cores of Ca {\sc ii}  H \& K become bright in emission when overlying magnetic heating elevates chromospheric temperatures \citep{Linsky:1980}. This activity is linked to the same magnetically-driven processes that give rise to radio emission, such as gyro-synchrotron radiation from accelerated electrons in stellar coronae \citep{Gudel:2002}. Therefore, Ca {\sc ii}  H \& K observations provide complementary optical tracers of the magnetic fields and heating processes that also power radio outbursts and quiescent emission in active stars \citep{Giampapa:1989}.

    The diagnostic relevance of Ca {\sc ii}  H \& K varies across different classes of radio-emitting stars. In cool dwarfs (e.g., M dwarfs), enhanced Ca {\sc ii} emission is a hallmark of strong magnetic activity that often correlates with intense radio bursts, although these bursts are not always simultaneous with the optical activity \citep{Slee:2003, Reiners:2008, Mlangeni:2026}. In RS CVn binaries and FK Comae stars, strong, persistent Ca {\sc ii} H \& K emission accompanies large-scale magnetic structures that also fuel quiescent radio emission \citep{Drake:1989}. Conversely, in early-type stars (e.g., O and B stars), which may exhibit radio emission via thermal winds or magnetospheres, Ca {\sc ii}  H\&K are typically weak or absent, reflecting the lack of convectively driven  dynamos.

    In an attempt to expand the known number of radio stars in our Milky Way, \cite{Egbo:2025} searched for radio-emitting stars by cross-matching the \cite{Mutale:2025} compact source catalogue derived from \cite{Goedhart:2024} South African Radio Astronomical Observatory (SARAO) MeerKAT Galactic Plane Survey (SMGPS) with two X-ray surveys, \textit{ROSAT} and \textit{eROSITA}. Both X-ray surveys cover the soft X-ray band between $0.1 - 2.4 \ \mathrm{keV}$ \citep{Boller:2016, Merloni:2024}. The X-ray properties of the active stars derived from the cross-match have been reported in \cite{Egbo:2026}. This paper reports the optical spectroscopic follow-up performed using facilities at the South African Astronomical Observatory (SAAO). The objectives of this study include the following: to obtain the spectroscopic information covering the blue- to red-portion of the optical window and search for emission lines (such as Ca {\sc ii} H\&K, H$\alpha$ and the Ca {\sc ii} infrared triplets) for characterising these chromospherically active stars.


    \section{Data Description} \label{sec:data}
    \subsection{Source Identification}

    Following the South African Radio Astronomical Observatory (SARAO) MeerKAT Galactic Plane Survey (SMGPS; \citealt{Goedhart:2024}), a radio-optical crossmatch analysis resulted in hundreds of candidate stellar source counterparts \citep{Egbo:2025}. The identified stars include early-type massive stars, such as Wolf-Rayet and OB-type stars; main-sequence single and binary stars; as well as red dwarfs. A cross-match of this radio--optical sample with X-ray catalogs from the \textit{eROSITA} all-sky survey (eRASS1; \citealt{Predehl:2021, Merloni:2024}) and the \textit{ROSAT} all-sky survey (RASS; \citealt{Truemper:1982, Boller:2016}) resulted in 137 radio--optical--X-ray sources. We established that the SMGPS specific radio luminosity and soft X-ray luminosity follow the G\"{u}del-Benz relation for active stars, although the classic relation accounts for the upper-limit of the population. In addition, we found that the {\it eROSITA} sensitivity allowed us to find more luminous early-type stars whose X-ray emission is not a result of coronal activity, but rather a result of other processes such as wind emission \citep{Egbo:2026}. 

    For the optical spectroscopic follow-up, we applied a set of selection criteria based on the optical properties of the active stars in the sample of radio--X-ray stars from \cite{Egbo:2026}. Our optical selection is drawn from \textit{Gaia} DR3 \citep{GaiaCollaboration:2023}. \textit{Gaia} obtained low-resolution spectra for stars with $G \leq 17.6$ mag, and we made use of the astrophysical parameters provided by the \textit{Gaia} Data Processing and Analysis Consortium (DPAC; \citealt{GaiaCollaboration:2023}). From the radio–X-ray sample presented in \cite{Egbo:2026}, we identified 99 stars with spectral types G, K, and M. These classifications are based on the \textit{Gaia} DR3 Extended Stellar Parametrizer for Emission-Line Stars (ESP-ELS) which classifies stars based on spectral types using \textit{Gaia} BP and RP low-resolution spectra for stars \citep{Fouesneau:2023}.

    To assess emission activity, we examined the H$\alpha$ equivalent width measurements from the \textit{Gaia} DR3 low-resolution spectra. We found that 40 of the 99 stars show H$\alpha$ in emission, with $\mathrm{H}\alpha_{\mathrm{EW}} < -0.1$ \AA. While this confirms clear activity in a subset of the sample, the remaining stars  do not show detectable H$\alpha$ emission at the \textit{Gaia} spectral resolution and sensitivity. Given that chromospheric activity can be variable and that emission may be missed in low-resolution spectra, we retained all 99 stars for optical spectroscopic follow-up. These observations were carried out using multiple spectrographs at the South African Astronomical Observatory (SAAO).

    \subsection{Spectroscopic Data Collection}\label{telescope}

    We began by using the {\it Lesedi} robotic telescope (described fully in \ref{mookodi}) to observe the majority of our targets, obtaining initial spectra and assessing their H$\alpha$ spectral properties (the {\it Mookodi} spectrograph on {\it Lesedi} has a blue limit of 4000 \AA, so Ca II H \& K lines are not observable). A large fraction of these stars displayed H$\alpha$ emission. This motivated a follow-up spectroscopic campaign with SpUpNIC (see \ref{spupsect}) and SALT HRS (see \ref{saltsect}). The SpUpNIC observations were aimed at investigating the Ca {\sc ii} H\&K region as a diagnostic of chromospheric activity for the late-type star candidates, whereas the SALT HRS observations were taken to obtain high-resolution spectra for a subset of the sample, given that they cover a longer bandwidth and the resolution is strong enough to resolve the Ca {\sc ii} IRT lines. As a result, the SpUpNIC and SALT HRS samples were not designed to be statistically representative of the full {\it Mookodi} sample but rather were chosen as follow-up targets based on evidence of activity in the initial observations, coupled with practical observing constraints such as target visibility and telescope time allocation.
    
    Exposure times were determined individually for each target based on its apparent brightness and the signal-to-noise ratio required for the scientific analysis. Also, the observations are heterogeneous and should not be interpreted as having uniform sensitivity across the {\it Mookodi}, SpUpNIC, and SALT HRS datasets.

    \subsubsection{Lesedi-{\it Mookodi}}\label{mookodi}
    
    Lesedi is a 1-m telescope operated by the South African Astronomical Observatory (SAAO) at the Sutherland site and used for transient, time-domain and target-of-opportunity observations as part of the Intelligent Observatory (‘IO’) system at the Observatory \citep{Worters:2016, Potter:2021, Erasmus:2025}. \textit{Mookodi}, one of the primary instruments on Lesedi, is a low-resolution optical spectrograph and imager covering the visible wavelength range of $\sim$400--800~nm \citep{Erasmus:2024}. {\it Mookodi} is based on the SPRAT design developed at Liverpool John Moores University and offers resolution powers of $R \simeq 300$, enabling rapid spectroscopic classification and follow-up of transient and variable sources \citep{Piascik:2018, Erasmus:2024}. Our {\it Mookodi} observations were primarily aimed at detecting H$\alpha$ emission. The spectra were reduced using the {\tt ASPIRED\footnote{{\tt ASPIRED} is a Python data reduction routine for optical image and spectral data reduction. For more details, see \url{https://aspired.readthedocs.io}.}} spectroscopic reduction package \citep{Lam:2023}. The information about the {\it Mookodi} observations used in this work is shown in Appendix \ref{mkdlog}.

    \subsubsection{1.9 m SpUpNIC}\label{spupsect}
    We obtained spectroscopic observations using the SpUpNIC (Spectrograph Upgrade: Newly Improved Cassegrain) spectrograph on the SAAO 1.9\,m (74-inch) telescope \citep{Crause:2016, Crause:2019}. The instrument offers a choice of low- to medium-resolution gratings covering different wavelength ranges. Two gratings were utilised for this work and are described below:

    \begin{itemize}
        
        \item SpUpNIC medium-resolution blue configuration (1200 l/mm): This grating offers a medium resolution with a dispersion of $1.3\,$ \AA \, $\rm pixel^{-1}$ and a resolving power of $R \sim 2500$ \citep{Crause:2016}. It covers the wavelength range 3800 to 4900 \AA. This wavelength range provides the opportunity to search for the Ca {\sc ii} H\&K emission lines, which are the primary lines for chromospheric identification. Information about the targets observed with this configuration is given in Appendix \ref{gr4log}.

        \item SpUpNIC low-resolution wideband configuration (300 l/mm): This low-resolution grating has a dispersion corresponding to a resolving power of $R \sim 700$ and covers a wavelength range of $3600 - 8000$ \AA\, \citep{Crause:2016}. This allows us to investigate the hydrogen Balmer lines, including $\rm H\alpha$, in addition to the Ca {\sc ii} H\&K lines. In Appendix \ref{gr7log}, we show the observation dates and exposure times for the observed stars. 
    \end{itemize}

    \noindent
    The calibration data (Cu-Ar arc, bias and flat field exposures) were collected at regular intervals each night for bias subtraction, flat fielding and wavelength calibration. Data reductions were carried out using the SpUPNIC data reduction utilities, which follow the standard {\tt IRAF} routines.

    \subsubsection{SALT High Resolution Spectrograph (HRS)}\label{saltsect}
    The High Resolution Spectrograph (HRS; \citealt{Barnes:2008, Bramall:2010, Bramall:2012, Crause:2014}) is an \'echelle spectrograph on the Southern African Large Telescope (SALT; \citealt{Buckley:2006, O'Donoghue:2006}) in Sutherland, South Africa. It comprises dual red and blue channels covering the region $3700-8900$ \AA\, with a resolution $(\rm R \sim 15000)$. SALT HRS spectra are therefore capable of detecting all of the chromospheric lines (Ca {\sc ii} H \& K, H$\alpha$ and the Ca {\sc ii} IRT. The primary reduction of the HRS spectroscopy was conducted using the SALT science pipeline\footnote{SALT HRS pipeline can be accessed at \url{https://astronomers.salt.ac.za/software/hrs-pipeline}} \citep{Crawford:2010}, which includes overscan correction, bias subtraction, and gain correction. The rest of the reduction was done using the MIDAS HRS pipeline described in detail in \cite{Kniazev:2016, Kniazev:2019}. The information about the HRS observations is shown in Appendix \ref{hrslog}.

    \section{Spectral Analysis Methodology} \label{sec:method}
    \subsection{Chromospheric Activity}

    \subsubsection{Equivalent Width Measurements}
    We investigated the $\rm H\alpha$ emission indicator of our sample. In the case where the stars show either emission or absorption lines, we normalised their spectra and calculated the Equivalent Width (EW) for each identified line using the relation: 

    \begin{eqnarray}
        &&{\rm EW}\,=\,{\int }_{{\lambda }_{1}}^{{\lambda }_{2}}\left(1-\frac{{F}_{I\lambda }}{{F}_{C\lambda }}\right)d\lambda ,
    \end{eqnarray}

    \noindent
    where $F_{C\lambda}$ is the local continuum flux near the line region, and $F_{I\lambda}$ is the spectral flux over the line area \citep{West:2004, West:2011, Newton:2017, Zhang:2025}. In the case of Ca {\sc ii} H\&K, these ranges are $3925 - 3940$ \AA\, and $3960 - 3975$ \AA\, for the K and H line respectively.

    \subsection{Activity Indices}
    \subsubsection{S-index}
    The Mount Wilson $S$-index quantifies stellar chromospheric activity by comparing the flux in the cores of the Ca {\sc ii} H and K emission lines to the flux in the adjacent continuum \citep{Wilson:1968, Vaughan:1978, Vaughan:1980, Duncan:1991, Cincunegui:2007, IbanezBustos:2023}. The parameter represents the total Ca {\sc ii} H \& K line emission, including a possible photospheric component that does not arise from magnetic activity \citep{Han:2025}. The Mt Wilson $S-$index is defined as:
    
        \begin{eqnarray}
            S_{\rm MW} = \alpha\frac{F_H + F_K}{F_R + F_V}
        \end{eqnarray}
    
    \noindent
    where $F_H$ and $F_K$ are the fluxes in two triangular bandpasses centred on the cores of the Ca {\sc ii} H\&K lines ($3968.46 $ \AA \, and $3933.66)$ \AA \, with widths of $2.18$ \AA \, (at the base), whereas $F_R$ and $F_V$ are the estimates of the pseudo-continuum fluxes in two rectangular $20$ \AA \, bandpasses centred on the continuum on either side of the Ca {\sc ii} H\&K lines at $3901.07$ \AA \, and $4001.07$ \AA \, respectively \citep{BoroSaikia:2018, deGrijs:2021, Marvin:2023, Yu:2025}. $\alpha$ is a normalisation constant, linking any sample of stars detected with other spectrographs to the Mt Wilson instrument \citep{Duncan:1991, Mittag:2013, deGrijs:2021}. Since our analysis does not require calibration onto this system, we adopt $\alpha=1$, effectively working with an instrumental S-index (e.g. \citealt{Han:2025}).

        \begin{eqnarray}
            S = \frac{F_H + F_K}{F_R + F_V}
        \end{eqnarray}

    \subsubsection{Chromospheric Emission Ratio}

    Chromospheric activity has traditionally been characterised by the $R_{\rm HK}^{\prime}$ index, defined as the ratio of the emission in the core of the Ca {\sc ii} H \& K lines to the total bolometric emission of the star. This index can be expressed as follows (e.g. \cite{Gondoin:2018}):
    
        \begin{eqnarray}
            R_{\mathrm{HK}}^{\prime} = \frac{F_H^{\prime} + F_K^{\prime}}{\sigma T_{\rm eff}^4}
            \label{rhk}
        \end{eqnarray}

    \noindent
    where $F_H^{\prime}$ and $F_K^{\prime}$ are the emission fluxes in the cores of the Ca {\sc ii} H \& K lines, $\sigma$ is the Stefan–Boltzmann constant, and $T_{\rm eff}$ is the effective surface temperature \citep{Gondoin:2020}. 

    The Ca {\sc ii} H \& K fluxes were measured by integrating the line profiles following continuum normalisation, and the associated uncertainties were estimated by propagating the spectral noise together with the uncertainty in continuum placement and line integration. The same uncertainty estimation procedure was applied consistently to all reported measurements.


\section{Results \& Discussion}

    \subsection{Chromospheric activity indicators}\label{chactivity}

    From our data reduction and analysis of all the optical spectra across different telescopes described in section \ref{telescope}, we identified 46 unique stars showing chromospheric activity of either Ca {\sc ii} H\&K, IRT or both.
    Among these, 39 stars observed with the SpUpNIC medium-resolution blue configuration display Ca {\sc ii} H and K emission features, whereas 18  observed with SALT HRS, show Ca {\sc ii} IRT. Of these 46 stars, 23 were previously known to be chromospherically active owing to their detection in \cite{Freund:2025}. The spectra of these stars are shown in Figs. \ref{chmofig}, \ref{hrsalpha} \& \ref{hrscairt}. For each target, we computed the \textit{S}-index, the chromospheric activity index $R_{\mathrm{HK}}^{\prime}$, and other line parameters such as the full-width at half maximum (FWHM) and equivalent width (EW). The $S$-index values span from $0.55$--$7.36$. When converted to $R_{\rm HK}^{\prime}$ as defined in Eq.~\ref{rhk}, the activity levels span $\log R_{\rm HK}^{\prime} \sim -6.4$ to $-2.8$. According to \cite{Henry:1996, Zhang:2024}, stellar chromospheric activity is classified into four levels: very active (log $R_{\rm HK}^{\prime} > -4.20$), active (log $R_{\rm HK}^{\prime}$ from $-4.75$ to $-4.20$), inactive (log $R_{\rm HK}^{\prime}$ from $-5.10$ to $-4.75$), and very inactive (log $R_{\rm HK}^{\prime} < -5.10$). Most stars in our sample lie above log $R_{\rm HK}^{\prime} = -4.75$, corresponding to the active and very active levels. The estimates are shown in Table \ref{ewgr4}, including the $G_{\rm BP}-G_{\rm RP}$ colour, radio luminosity ($R_L$), and X-ray luminosity ($X_L$).

    \begin{table*}
        \centering
        \caption{Activity levels and radio, optical, and X-ray properties of the chromospheric active stars}
        \begin{threeparttable}
        \begin{tabular}{lrclrrccc}
    
    \toprule
    
    \parbox{1.3cm}{Target Name} & 
    \parbox{0.7cm}{$G$ \\ (mag)} &  
    \parbox{1.5cm}{ $G_{\rm BP}-G_{\rm RP}$ \\ (mag)} &  
    \parbox{1.0cm}{$P_{\rm rot}$ \\ (days)} & \parbox{0.7cm}{Dist \\ (pc) } & 
    \parbox{0.9cm}{$\log R_{\rm HK}^{\prime}$ \\ \textemdash} & 
    \parbox{1.4cm}{$\log R_L$ \\ $(\mathrm{erg\,s^{-1}\,Hz^{-1}})$} & 
    \parbox{1.2cm}{$\log X_L$ \\ $(\mathrm{erg\,s^{-1}})$} & 
    \parbox{1.0cm}{$\log L_{\rm bol}$ \\ $(\mathrm{erg\,s^{-1}})$} \\

    \midrule
    ATO J123.1111--34.1401\tnote{(a)} & 14.48 & 1.49 & 5.48 \tnote{(1)} & 415.9 & --5.71 & 16.32 & \textemdash & 33.09 \\
    OGLE GD-CEP-177\tnote{(a)} & 12.62 & 1.96 & 1.82\tnote{(2)} & 358.5 & --3.71 & 16.84 & 31.30 & 34.02 \\
    Gaia DR3 5542859804655658880\tnote{(a)} & 11.50 & 0.93 & \textemdash & 215.2 & --3.93 & 15.91 & 30.42 & 33.59 \\
    Gaia DR3 5528151538832694144\tnote{(a)} & 15.20 & 1.79 & 0.89\tnote{(2)} & 284.3 & --4.44 & 15.92 & \textemdash & 32.47 \\
    Gaia DR3 5527871021626231424\tnote{(a)} & 11.86 & 1.10 & 3.02\tnote{(2)} & 349.7 & --4.26 & 16.10 & 30.76 & 33.98 \\
    TYC 7665-2262-1\tnote{(a,b)} & 12.04 & 1.27 & \textemdash & 475.4 & --4.08 & 16.77 & 30.84 & 34.18 \\
    L 459-76\tnote{(a)} & 11.13 & 1.87 & 2.02\tnote{(3)} & 43.6 & --4.54 & 15.00 & 28.21 & 32.47 \\
    Gaia DR3 5527703109886064768\tnote{(a)} & 15.64 & 2.79 & \textemdash & 336.2 & --3.18 & 16.77 & \textemdash & 32.75 \\
    Gaia DR3 5523337086650972416\tnote{(a)} & 15.31 & 2.82 & 0.64\tnote{(4)} & 316.3 & --4.38 & 16.64 & \textemdash & 32.84 \\
    CD-45 4744\tnote{(a)} & 8.65 & 1.36 & 1.36\tnote{(2)} & 547.9 & --4.53 & 17.16 & 31.72 & 35.66 \\
    TYC 8156-1758-1\tnote{(a,b)} & 10.10 & 1.48 & 2.57\tnote{(3)} & 174.8 & --4.05 & 16.56 & 31.06 & 34.08 \\
    TYC 8169-1002-1\tnote{(a)} & 11.78 & 1.66 & 1.17\tnote{(3)} & 96.2 & --4.39 & 15.02 & 29.62 & 32.89 \\
    GSC 08584-01898\tnote{(a,b)} & 11.97 & 1.36 & \textemdash & 133.7 & --4.45 & 15.24 & 29.92 & 33.10 \\
    HD 82533\tnote{(a)} & 8.09 & 0.92 & 0.49\tnote{(3)} & 54.4 & --4.05 & 15.61 & 30.43 & 33.75 \\
    UCAC4 174-040115\tnote{(a,b)} & 13.83 & 2.52 & 1.67\tnote{(2)} & 138.9 & --4.38 & 15.29 & 29.48 & 32.71 \\
    Gaia DR3 5307640461803223040\tnote{(a,b)} & 12.95 & 1.61 & \textemdash & 1242.1 & --4.62 & 17.73 & 31.32 & 34.65 \\
    HD 87525\tnote{(a)} & 7.66 & 1.50 & 5.44\tnote{(3)} & 326.6 & --4.67 & 16.51 & 31.65 & 35.60 \\
    2MASS J10462256--6113379\tnote{(a,b)} & 11.27 & 1.78 & 1.70\tnote{(5)} & 97.6 & --3.42 & 15.53 & 30.10 & 33.11 \\
    UCAC4 146--088848\tnote{(a,b)} & 13.40 & 2.66 & 4.24\tnote{(6)} & 87.9 & --2.93 & 15.63 & 29.26 & 32.49 \\
    UCAC4 143-089465\tnote{(a)} & 11.77 & 1.45 & \textemdash & 1267.1 & --2.79 & 17.06 & 30.64 & 35.14 \\
    ASAS J115948--6136.2\tnote{(a)} & 11.41 & 1.75 & 3.04\tnote{(3)} & 113.3 & --4.88 & 15.04 & 30.09 & 33.18 \\
    HD 104919\tnote{(a)} & 9.24 & 0.90 & 2.61\tnote{(3)} & 100.9 & --4.21 & 15.52 & 30.56 & 33.83 \\
    HD 106506\tnote{(a)} & 8.28 & 0.81 & 2.88\tnote{(3)} & 106.9 & --4.17 & 16.00 & 30.78 & 34.26 \\
    Gaia DR3 6057905002869874816\tnote{(a)} & 12.01 & 1.57 & 5.26\tnote{(3)} & 1612.4 & --4.96 & 17.49 & 31.34 & 35.25 \\
    FP Cru\tnote{(a,b)} & 10.61 & 1.55 & 4.34\tnote{(3)} & 100.3 & --4.30 & 16.18 & 30.35 & 33.40 \\
    2MASS J12271665-6239142\tnote{(a)} & 10.49 & 1.31 & 3.95\tnote{(3)} & 103.7 & --4.74 & 15.19 & 30.03 & 33.47 \\
    CD-62 657\tnote{(a)} & 9.12 & 1.00 & 3.37\tnote{(2,3)} & 107.3 & --4.24 & 15.84 & 30.63 & 34.05 \\
    CPD-63 2367\tnote{(a)} & 9.73 & 1.27 & 0.71\tnote{(2,3)} & 104.1 & --4.35 & 15.73 & 30.39 & 33.78 \\
    FI Cru\tnote{(a)} & 10.34 & 1.35 & 1.28\tnote{(2,5,6)} & 111.7 & --4.53 & 16.24 & 30.39 & 33.60 \\
    HD 112245\tnote{(a)} & 9.16 & 1.00 & 1.25\tnote{(2,3,5,6)} & 77.7 & --4.11 & 15.71 & 30.56 & 33.76 \\
    Gaia DR3 5863410879175831168\tnote{(a,b)} & 12.11 & 1.99 & 3.04\tnote{(2)} & 95.4 & --4.47 & 15.15 & 29.70 & 33.07 \\
    HD 312273\tnote{(a)} & 9.94 & 1.12 & 1.20\tnote{(7)} & 117.1 & --4.18 & 15.46 & 30.31 & 33.80 \\
    V755 Cen\tnote{(a,b)} & 11.58 & 1.74 & 0.98\tnote{(3)} & 115.0& --4.09 & 15.86 & 29.98 & 33.13 \\
    V851 Cen\tnote{(a)} & 7.35 & 1.34 & 5.58\tnote{(3)} & 73.9 & --4.27 & 15.69 & 30.72 & 34.43 \\
    Gaia DR3 5867188934252852224\tnote{(a,b)} & 12.70 & 1.84 & 5.66\tnote{(3)} & 2002.4 & --4.02 & 18.32 & 31.71 & 35.16 \\
    CD-59 5220\tnote{(a)} & 10.71 & 1.34 & 5.74\tnote{(3)} & 111.9 & --4.93 & 15.71 & 30.45 & 33.45 \\
    V841 Cen\tnote{(a)} & 8.38 & 1.39 & 5.98\tnote{(3,8)} & 94.9 & --4.15 & 16.69 & 31.06 & 34.24 \\
    Gaia DR3 5883028464399762048\tnote{(a)} & 12.05 & 1.52 & 1.99\tnote{(3)} & 776.1 & --4.47 & 17.32 & 31.40 & 34.60 \\
    Gaia DR3 5935656077388451712\tnote{(a)} & 12.57 & 2.20 & \textemdash & 1107.6 & --6.40 & 17.35 & 31.34 & 34.70 \\
    
    Gaia DR3 6053784067641276160\tnote{(b)} & 12.11 & 1.94 & \textemdash & 113.6 & \textemdash & 15.93 & 29.96 & 33.23 \\
    UCAC4 134-060461\tnote{(b)} & 13.14 &  2.53 & \textemdash & 104.3 & \textemdash & 15.69 & 29.44 & 32.74 \\
    BK Cru\tnote{(b)} & 13.08 & 2.58 & \textemdash & 104.2 & \textemdash & 15.68 & 29.37 & 32.76 \\
    Gaia DR3 6054608701362026112\tnote{(b)} & 12.59 & 2.54 & \textemdash & 50.0 & \textemdash & 14.98 & 29.18 & 32.32 \\
    2MASS J17275960-3607344\tnote{(b)} & 11.57 & 1.56 & \textemdash & 129.1 & \textemdash & 15.77 & 29.96 & 33.23 \\

    Gaia DR3 4156018236362660352\tnote{(b)} & 13.53 & \textemdash & \textemdash & 221.4 & \textemdash & 16.30 & \textemdash & \textemdash \\
    Gaia DR3 5883899483767476224\tnote{(b)} & 13.73 & \textemdash & \textemdash & 2570.6 & \textemdash & 19.18 & \textemdash & \textemdash \\
    
    \bottomrule
    \end{tabular}

    \begin{tablenotes}
        \item[(a)] Ca {\sc ii} H\&K emission detected.
        \item[(b)] Ca  {\sc ii} Infrared triplet (IRT) emission detected.
        \item \textbf{References:} (1) \cite{Jayasinghe:2018}; (2) \cite{Watson:2006}; (3) \cite{Green:2023}; (4) \cite{Kounkel:2022}; (5) \cite{Colman:2024}; (6) \cite{Lin:2024}; (7) \cite{Vach:2024}; (8) \cite{Samus':2017}.
    \end{tablenotes}
    \label{ewgr4}
    \end{threeparttable}
    \end{table*}

    \begin{figure}
        \centering
        \includegraphics[width=\columnwidth]{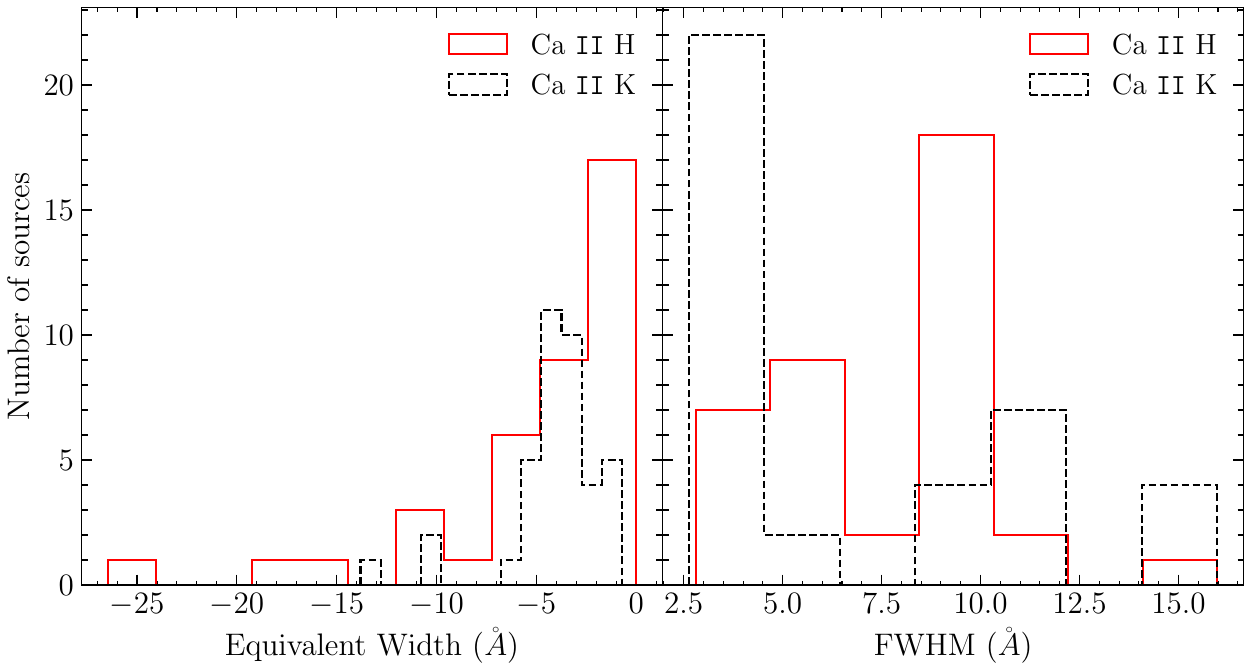}
        \caption{Distribution of equivalent widths (EW) and full width at half maximum (FWHM) for the Ca {\sc ii} H\&K lines obtained with SpUpNIC medium-resolution blue configuration.}
        \label{ew_fwhm}
    \end{figure}

    \subsubsection{Line Diagnostics of the Ca {\sc ii} H\&K Lines}\label{lindg}

        We measured the EW and FWHM of the Ca {\sc ii} H and K emission cores for all 39 stars observed with the SpUpNIC medium-resolution blue configuration. The EWs span -13.1 \AA \, $ \leq \mathrm{EW_H} \leq -0.5$\,\AA\ for Ca {\sc ii} H, and -21.6 \AA \, $ \leq \mathrm{EW_K} \leq -0.2$\,\AA\ for Ca {\sc ii} K. The corresponding FWHM values range from $2.6$ \, \AA \, to $15.9$\,\AA\ for both lines, although a few stars show markedly broader cores. Such broadening may be attributed to rapid rotation or unresolved binarity, both of which can enhance chromospheric line widths through Doppler and magnetic effects \citep{Noyes:1984, Strassmeier:1990, Schroder:2009}. The distributions of EW and FWHM are presented in Fig. \ref{ew_fwhm}.
        
        The variation in EW among our targets reflects a broad range of chromospheric activity levels. Stars with larger (more negative) EW values exhibit stronger core emission, consistent with enhanced magnetic heating in the lower chromosphere \citep{Linsky:1979, Schroder:2009}. The observed spread in FWHM suggests differences in line broadening mechanisms that could arise from stellar rotation, microturbulence, or plage coverage \citep{Lopez-Santiago:2010, Reiners:2012}. As the observations were taken at random phases, some of the variation may reflect rotational or activity phase differences between stars.

        \begin{figure}
            \centering
            \includegraphics[width=\columnwidth]{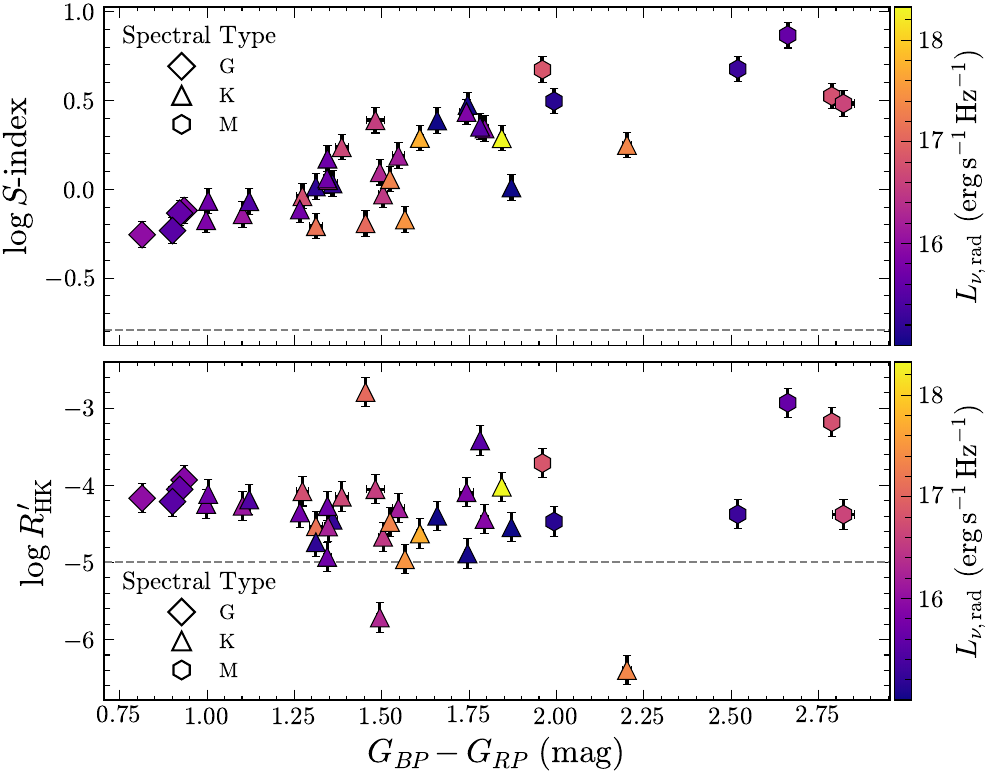}
            \caption{Activity plot for 39 chromospherically active stars. The $S-$index and activity index, $R_{\rm HK}^{\prime}$ as a function of $G_{\rm BP}-G_{\rm RP}$ are shown in the top and bottom panels. Each is colour-coded with the log of the radio luminosity. In addition, the solar minimum $S-$index, and $R_{\mathrm{HK}}^{\prime}$ is shown in both plots as dashed horizontal lines, and the spectral classes are from \protect\cite{Fouesneau:2023}.}
            \label{bvsi}
        \end{figure}

        \subsubsection{Activity as a Function of Stellar Colour}\label{rhkcol}

            Chromospheric activity is known to vary systematically with stellar spectral type, particularly among late-type main-sequence stars (G, K, and M), owing to differences in convective zone depth and dynamo efficiency \citep{Noyes:1984, Mamajek:2008}. Also, chromospheric activity may be influenced by intrinsic stellar variability and flare events, which can contribute to the observed scatter in the activity indices. Previous works in literature used $B-V$ as a proxy for effective temperature and spectral type. In our case, we used the {\it Gaia} $G_{\rm BP}-G_{\rm RP}$ colour index as the photometric proxy for stellar effective temperature and spectral type throughout this analysis. The {\it Gaia} photometry provides a homogeneous and consistent dataset for our  sample and provides a more reliable comparison of chromospheric activity for our case. The spectral classifications are adopted from \cite{Fouesneau:2023} and are consistent with the observed distribution of $G_{\rm BP}-G_{\rm RP}$ colours shown in Fig. \ref{bvsi}.

            The chromospheric activity indicator, $R_{\rm HK}^{\prime}$ spans approximately three orders of magnitude across the sample, with no clear monotonic dependence on the Gaia $G_{\rm BP}-G_{\rm RP}$ colour index. Instead, substantial scatter is present among stars of similar colour, indicating that chromospheric activity is not determined solely by effective temperature or spectral type \citep{Chahal:2023}. This behaviour is expected, as stellar activity is also influenced by factors such as age and rotation \citep{West:2015, Astudillo-Defru:2017}. Furthermore, late-type active stars are intrinsically variable, and chromospheric diagnostics such as Ca {\sc ii} H\&K and H$\alpha$ may vary on rotational, activity-cycle, and flare timescales. Since the activity indices presented here are derived from single-epoch observations, part of the observed scatter may reflect temporal variability rather than intrinsic differences between stars. The most chromospherically active objects in the sample are predominantly K- and M-type stars, with 37 of the 39 targets observed with the SpUpNIC medium-resolution blue configuration, exhibiting activity levels above the solar minimum value of $R_{\mathrm{HK (\odot)}}^{\prime} \approx -4.96$ \citep{Lockwood:2007}. Although the S-index generally increases towards redder colours (see top panel of \ref{bvsi}, this trend should be interpreted with caution because the S-index retains a dependence on the underlying photospheric contribution, making direct comparisons across spectral types less meaningful than those based on $R_{\rm HK}^{\prime}$ \citep{Mittag:2013}.
            
            There are two stars with high activity indices. {\it Gaia} DR3 5527703109886064768 is a young stellar object (YSO), and the {\it Gaia} ELS spectra classify it as an M dwarf \citep{Zari:2018, Fouesneau:2023}. UCAC4 143-089465 is a spectroscopic binary with a {\it Gaia} spectrum indicating it is a K-type star \citep{Fouesneau:2023, GaiaCollaboration:2023}. Another star showing strong activity is UCAC4 146-088848, a K-type YSO \citep{Udalski:2018, Zari:2018, He:2022, GaiaCollaboration:2023}. Below the solar $R_{\rm HK}^{\prime}$ minimum are two K stars: ATO J123.1111--34.1401, a variable star detected with ATLAS and Gaia DR3 5935656077388451712 \citep{Heinze:2018, GaiaCollaboration:2023}.

        \subsubsection{Chromospheric activity from the Ca {\sc ii} infrared triplet}

            In addition to the Ca\,{\sc ii} H\&K lines, chromospheric activity can be traced by the Ca\,{\sc ii} IRT at the wavelengths of $\lambda$\,8498, 8542, 8662\,\AA\, \citep{Andretta:2005, Linsky:2017}. These lines form in similar atmospheric layers as the H\&K lines, but are less affected by interstellar absorption, and provide higher signal-to-noise ratios in spectra of cool stars compared to the H\&K lines \citep{Linsky:1979, Busa:2007}. The \textit{Gaia} DR3 RVS spectra cover the Ca {\sc ii} IRT region (846–870 nm); however, no RVS spectra are available for our targets. 

            \begin{figure*}
                \includegraphics[width=\textwidth]{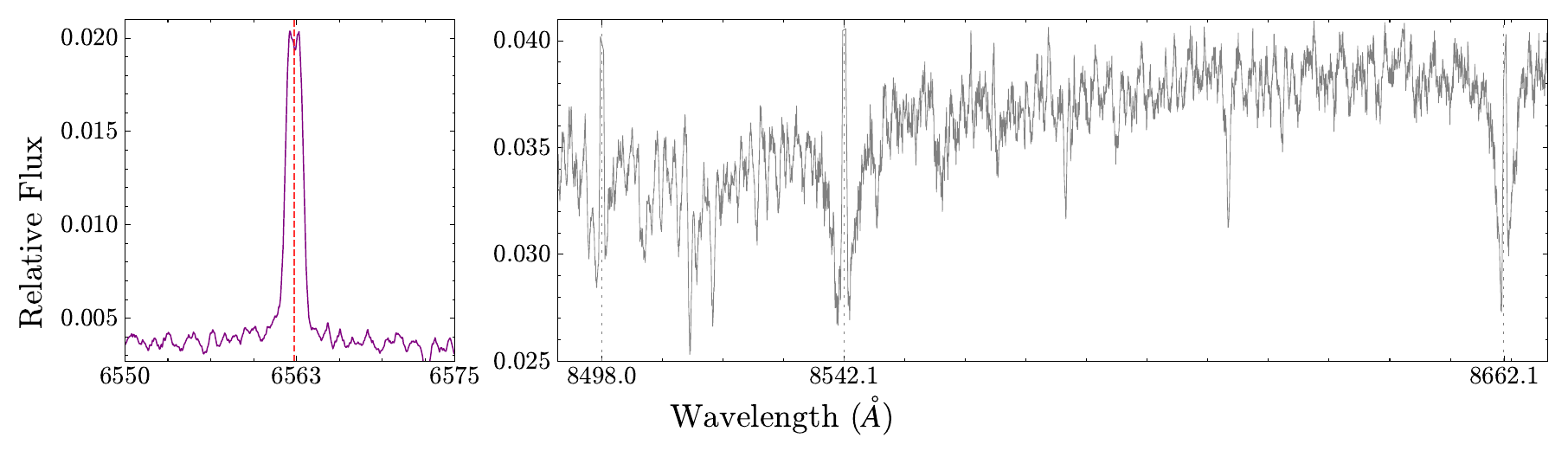}
                \caption{The SALT HRS spectra of UCAC4 146-088848 showing the emission lines of H$\alpha$ and  Ca {\sc ii} IRT lines.}
                \label{ucac2}
            \end{figure*}
            
            Using SALT HRS observations, we detected Ca {\sc ii} IRT lines in a subset of our sample, although the region is heavily contaminated by telluric lines. 
            In Fig. \ref{hrscairt}, we present the spectra of 18 of the 46 stars in our sample that show visible IRT lines, highlighting the emission cores that strengthen with increasing activity. Of these 18 stars, 11 also show the Ca {\sc ii} H\&K lines discussed earlier in Sect. \ref{chactivity}. An example of the SALT HRS spectra of UCAC4 146-088848 displaying H$\alpha$ and Ca {\sc ii} IRT is shown in Fig. \ref{ucac2}. We find that stars with higher $R_{\mathrm{HK}}^{\prime}$, such as UCAC4 146-088848 and {\it Gaia} DR3 5863410879175831168, also display enhanced IRT emission, confirming the consistency of both Ca {\sc ii} line diagnostics.

            However, in some cases, strong H\&K emission is not accompanied by proportionate IRT emission e.g. {\it Gaia} DR3 5867188934252852224, although a weak emission core is visible. This discrepancy may arise because the IRT lines, forming deeper in the chromosphere, are strongly influenced by the underlying photospheric continuum or basal flux, which can partially fill in their cores and mask emission \citep{Busa:2007, Linsky:2017}. 
            Alternatively, rapid stellar rotation may broaden the line profiles beyond the spectral window used for defining the activity index measurements, leading to apparently weaker IRT fluxes \citep{Martinez-Arnaiz:2010, Fuhrmeister:2022}. Furthermore, since the Ca {\sc ii} H\&K and IRT observations were obtained at different epochs, intrinsic variability in chromospheric activity may also contribute to the observed differences, with the IRT spectra likely representing a less active phase.

            \begin{figure*}
                \includegraphics[width=\textwidth]{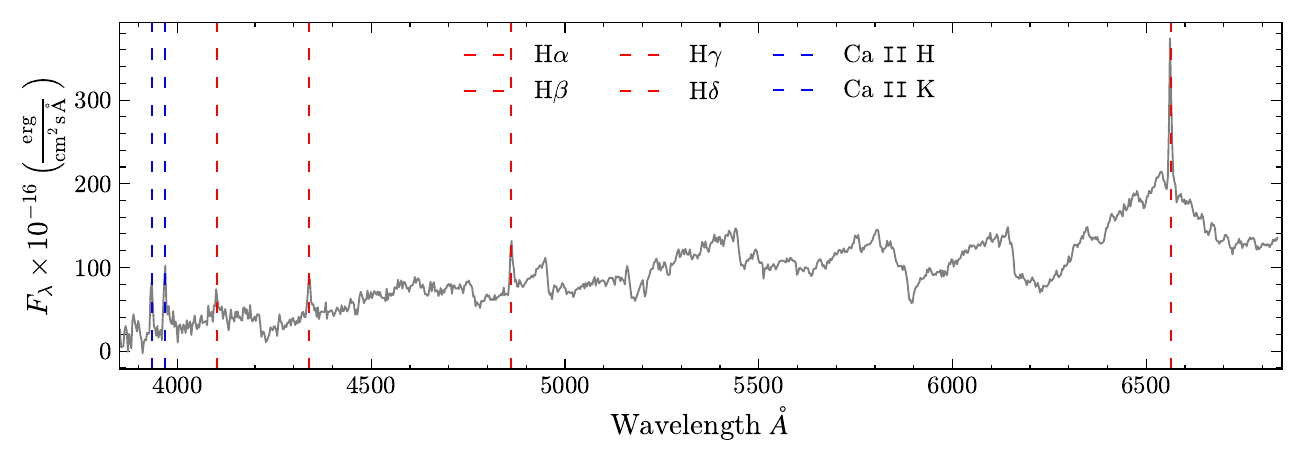}
                \caption{SpUpNIC low-resolution spectrum of UCAC4 146-088848 showing prominent hydrogen Balmer lines (H$\alpha$, H$\beta$, H$\gamma$, and H$\delta$) and the Ca\,{\sc ii} H \& K lines.}
                \label{ucac}
            \end{figure*}

        \subsubsection{H$\alpha$-emitting Stars}
            Alongside the Ca\,{\sc ii} lines, a subset of stars in our sample exhibit H$\alpha$ emission. The H$\alpha$ line at 6563\,\AA\ forms in the upper chromosphere, and its behaviour often differs from that of the Ca\,{\sc ii} H\&K diagnostics \citep{Linsky:2017}. In inactive stars, the line typically appears as a deep photospheric absorption feature, while in magnetically active stars, the core becomes increasingly filled in and, in strong cases, transitions into a pure emission profile above the continuum \citep{Cram:1979, Montes:1995, Cincunegui:2007, Linsky:2017}. Our spectra reveal distinct H$\alpha$ emission in several sources; in some cases, additional Balmer lines such as H$\beta$, H$\gamma$, and H$\delta$ are detected. Fig. \ref{hrsalpha} shows the non-flux-calibrated SALT HRS spectra of eighteen (18) stars from our sample exhibiting H$\alpha$ emission.  Morphologically, the profiles exhibit both narrow and broad emission features, with seven stars displaying clear double-peaked structures, typically associated with binary systems.

\begin{figure*}
                \includegraphics[width=\textwidth]{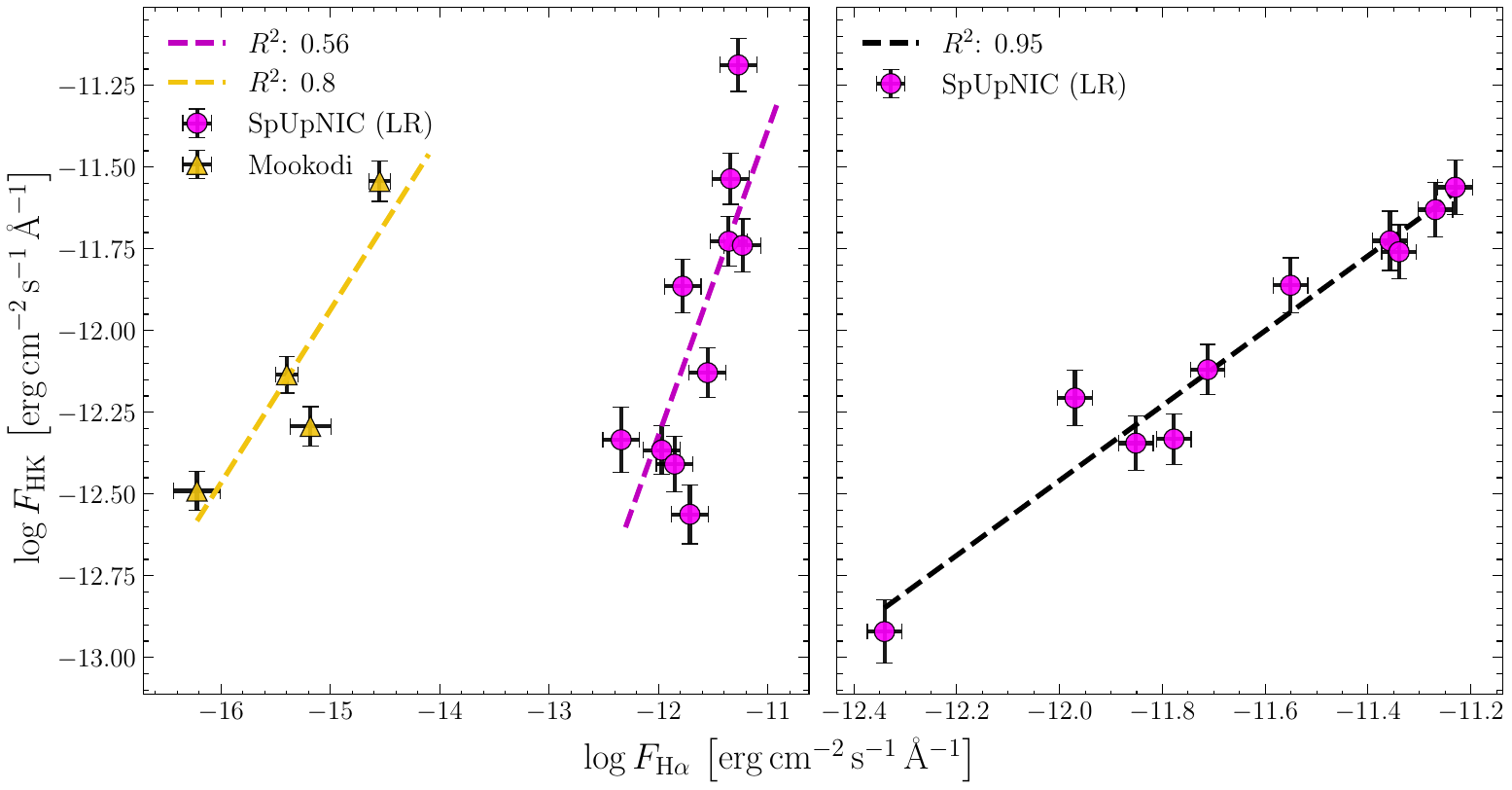}
                \caption{Average surface fluxes, $F_{\mathrm{HK}}$ versus $F_{\mathrm{H\alpha}}$. Left: H$\alpha$ fluxes from {\it Mookodi} (yellow points) and SpUpNIC low-resolution wideband configuration (magenta points) compared with Ca {\sc ii} H\&K fluxes from the SpUpNIC medium-resolution blue configuration; these measurements are not simultaneous. Right: Simultaneous H$\alpha$ and Ca {\sc ii} H\&K measurements obtained from the SpUpNIC low-resolution wideband configuration.}
                \label{hkha}
            \end{figure*}

            In addition, we compare the H$\alpha$ fluxes obtained from the low-resolution, flux-calibrated spectra from \textit{Mookodi} and the SpUpNIC low-resolution wideband configuration with the Ca {\sc ii} H\&K fluxes derived from the SpUpNIC medium-resolution blue configuration. A linear regression performed in log--log space reveals a correlation consistent with chromospheric activity trends (see the left panel of Fig. \ref{hkha}). The four stars represented by \textit{Mookodi} H$\alpha$ fluxes are included because they were not observed with the SpUpNIC low-resolution wideband configuration. 
            The non-simultaneous comparison between \textit{Mookodi} H$\alpha$ and SpUpNIC medium-resolution blue Ca {\sc ii} fluxes yields a strong correlation ($R^{2}\approx 0.80$). Likewise, the non-simultaneous SpUpNIC low-resolution wideband H$\alpha$ and medium-resolution blue Ca {\sc ii} observations show a moderate correlation ($R^{2}\approx 0.56$).    
            
            In contrast, the simultaneous SpUpNIC low-resolution wideband H$\alpha$ and medium-resolution blue Ca {\sc ii} observations exhibit the strongest correlation ($R^{2}\approx 0.95$), indicating that intrinsic stellar variability between observing epochs likely contributes to the increased scatter in the non-simultaneous cases. This behaviour is consistent with previously reported Ca {\sc ii}--H$\alpha$ flux correlations across late-type stars \citep{Cincunegui:2007, Walkowicz:2009, Santos:2010, GomesdaSilva:2011, IbanezBustos:2023}. In particular, \cite{IbanezBustos:2023} reported a positive correlation between Ca {\sc ii} and H$\alpha$ emission in dM stars, while \cite{Cincunegui:2007} found a similar trend for 109 FGKM-type stars when averaged over multiple epochs. However, comparisons based on individual observations of the same source reveal significant scatter, indicating temporal variability in chromospheric emission. Consistent with this, the left panel of Fig. \ref{hkha} shows a positive correlation between Ca {\sc ii} H\&K and H$\alpha$, although the lines were not observed simultaneously. 
            
            Despite the lower spectral resolution of the SpUpNIC low-resolution wideband configuration relative to the medium-resolution blue configuration, its broader wavelength coverage enables simultaneous monitoring of the Ca {\sc ii} H\&K and H$\alpha$ regions (e.g. Fig. \ref{ucac}, which shows the spectrum of UCAC4 146-088848). For stars exhibiting both Ca {\sc ii} H\&K and H$\alpha$ emission observed simultaneously with the low-resolution configuration, we also find a strong positive flux correlation ($R^{2} \approx 0.85$; see right panel of Fig. \ref{hkha}). 

            The \textit{Mookodi} and SpUpNIC observations were obtained at different epochs, in some cases separated by up to a year. Consequently, the observed differences in $F_{\mathrm{H}\alpha}$ likely reflect a combination of intrinsic stellar variability, including flare activity, together with differences in instrumental resolution and observing conditions.
            In general, the simultaneous presence of H$\alpha$ and Ca {\sc ii} H\&K emission in our sample is observed only among K and M dwarfs. As suggested by \cite{Meunier:2022}, enhanced fluxes in both diagnostics likely originate from active chromospheric plages, while overlying filaments can contribute to H$\alpha$ absorption.

            While this subsection focuses on stars exhibiting H$\alpha$ emission in our observations, it is also instructive to examine the behaviour of the full Ca {\sc ii} H\&K sample. Of the 39 stars observed with the SpUpNIC medium-resolution blue configuration, 25 showed neither H$\alpha$ emission nor absorption during our observing campaign. Using the H$\alpha$ equivalent width values from the {\it Gaia} DR3 astrophysical parameters table, we find that 19 of these 25 stars exhibited weak H$\alpha$ emission during the Gaia epoch. This suggests that the absence of H$\alpha$ emission in our data likely reflects temporal variability associated with stellar magnetic activity cycles.

    \subsection{Multiwavelength Relations}

        \subsubsection{X-ray to Bolometric Luminosity Ratio}
            Depending on the level of magnetic activity, active stars of late spectral types can also show X-ray emission from the magnetically heated coronae, particularly in the soft X-ray regime \citep{Gudel:2004, Robrade:2005, Reiners:2014, Linsky:2017, Fuhrmeister:2022, Freund:2024}. Previous studies have shown that stars with strong Ca {\sc ii} H\&K emission also tend to display enhanced coronal emission (e.g. \citealt{Schrijver:1983, Schrijver:1987, Mamajek:2008}). A direct comparison of chromospheric and coronal activity provides an important test of the connection between magnetic heating in the lower and upper stellar atmospheres \citep{Wright:2011, Mittag:2018, Fuhrmeister:2022, Freund:2025}. Of the 46 stars observed, only 39 show measurable Ca {\sc ii} H\&K emission and are therefore included in the chromospheric activity analysis based on the $S$-index and $R_{\rm HK}^{\prime}$. A subset of these stars has a corresponding X-ray detection.         
            The extended R\"oentgen Survey with an Imaging Telescope Array (eROSITA; \cite{Predehl:2021}) is an X-ray telescope onboard the Spectrum-R\"oentgen-Gamma mission (SRG; \cite{Sunyaev:2021}). The first eROSITA all-sky survey (eRASS1) detections, between December 2019 and June 2020, have been reported in \cite{Merloni:2024}, covering the Western Galactic Hemisphere. In addition, \cite{Freund:2024} using a Bayesian probabilistic approach in combination with {\it Gaia} DR3 photometric and astrometric information produced a sample of X-ray coronal-emitting stellar sources.

            To investigate the stellar chromospheric–corona connection, we used X-ray fluxes for 35 of the 39 stars observed with the SpUpNIC medium-resolution blue configuration, obtained from \cite{Merloni:2024} and \cite{Freund:2024}. Using distance estimates from \cite{Bailer-Jones:2021}, we computed the X-ray luminosities, $L_X$. Also, the bolometric\footnote{We computed $L_{\rm bol}$ from the de-reddened \textit{Gaia} $G$-band magnitude, applying bolometric corrections based on spectral class from a table based upon \cite{Pecaut:2013}, regularly updated at \url{http://www.pas.rochester.edu/~emamajek/EEM_dwarf_UBVIJHK_colors_Teff.txt}} luminosity, $L_{\mathrm{bol}}$ was calculated by first applying bolometric corrections to the de-reddened {\it Gaia}  $G$ mag. Using both luminosities, we derived the $L_{\rm X}/L_{\rm bol}$ ratios and compared them to the activity index, $R_{\rm HK}^{\prime}$. The plot of $L_{\rm X}/L_{\rm bol}$ against $R_{\rm HK}^{\prime}$ is shown in the left panel of Fig. \ref{lxlbol}. Thirty-three stars had $L_{\rm X}/L_{\rm bol} < -4$. Generally, the $L_{\rm X}/L_{\rm bol}$ for all the stars in this plot are generally within $L_{\rm X}/L_{\rm bol}> -5$, the saturation limit for coronally active stars. On the other hand, we established a positive correlation between $L_{\rm X}/L_{\rm bol}$ and $R_{\rm HK}^{\prime}$ with a linear fit of $\log (L_{\rm X}/L_{\rm bol}) = (0.50\pm 0.23) \log R_{\rm HK}^{\prime} -(1.18 \pm 0.97)$. \cite{Fuhrmeister:2022} established a similar correlation while studying the connection between the chromospheric Ca II H\&K emission and simultaneous X-ray measurements obtained from {\it eROSITA} for 183 dwarfs and derived a correlation between the X-ray to bolometric flux ratio and the activity index. However, we do not see a clear correlation between $R_{\rm HK}^{\prime}$ and the specific radio luminosity, $L_{\nu,R}$, just like in Fig. \ref{bvsi}. Notwithstanding this, we directly compare the radio and X-ray luminosities to investigate if they follow the canonical G\"{u}del–Benz relationship for active stars.

        \subsubsection{Radio and X-ray Luminosity}
            The radio and X-ray luminosities for active stars have been studied extensively for active stars. Initially, \cite{Guedel:1993, Benz:1994} established a linear relationship for the log of the radio and X-ray luminosities referred to as the canonical G\"{u}del–Benz relationship for active stars \citep{Guedel:1993, Benz:1994}, which states that the non-thermal gyrosynchrotron emission detected in the radio is correlated to the thermal coronal emission observed in the soft--X-ray wavelength \citep{Williams:2014, Driessen:2022, Vedantham:2022}. However, recent studies have shown that the G\"{u}del–Benz fit accounts for the upper limit of the active star samples \citep{Yiu:2024, Driessen:2024}. In our recent work, we also established this behaviour for stars using MeerKAT, {\it ROSAT}, and {\it eROSITA} data \citep{Egbo:2026}. Since the sample used in this work is a subset of the population, we attempted to investigate the G\"{u}del–Benz relation once more in comparison with the chromospheric activity index, $R_{\mathrm{HK}}^{\prime}$.

\begin{figure*}
                \centering
                \includegraphics[width=\textwidth]{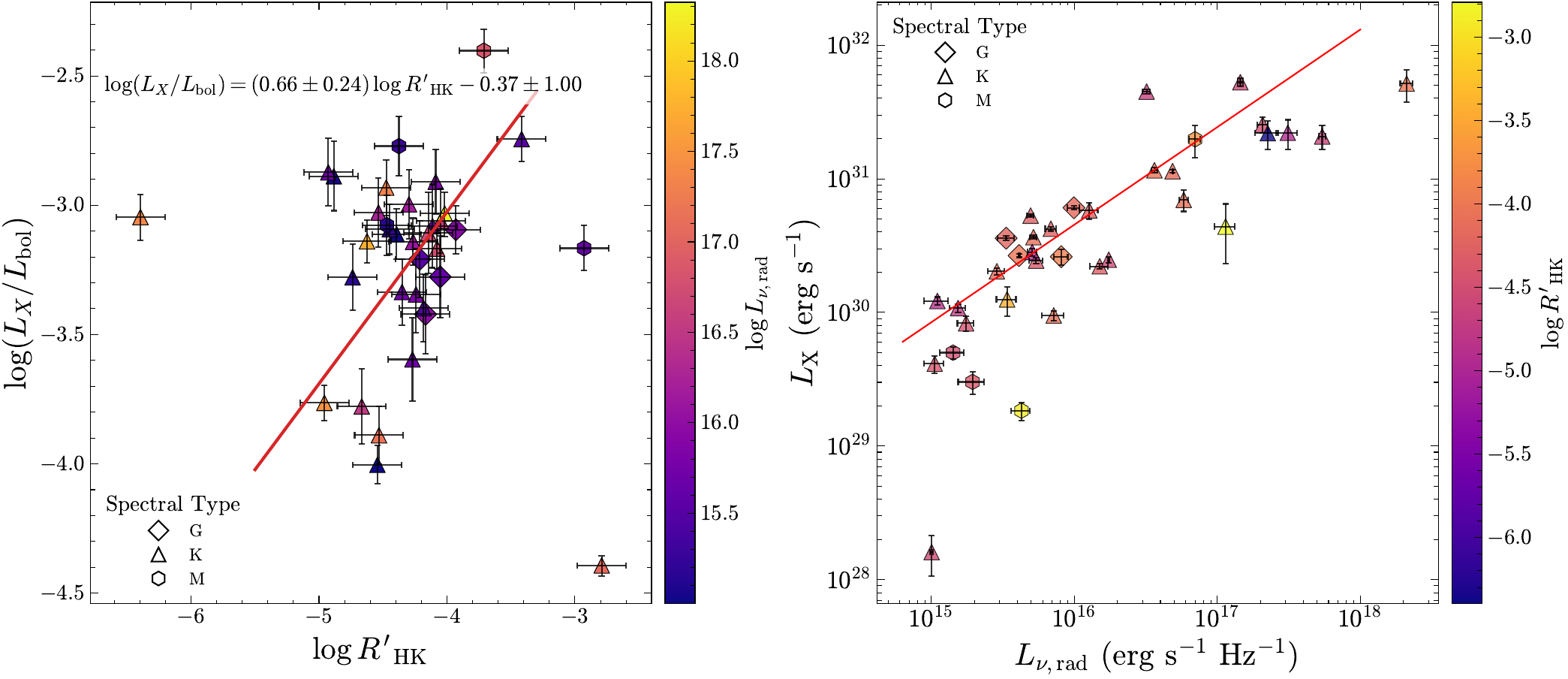}
                \caption{Left: X-ray to bolometric luminosity ratio, $L_{\rm X}/L_{\rm bol}$ as a function of the activity index,  $R_{\rm HK}^{\prime}$ for the 35 stars with X-ray fluxes.  Right: Their respective radio--X-ray luminosity plot. The solid line represents the G\"udel--Benz relation and is shown for comparison with our sample.}
                \label{lxlbol}
            \end{figure*}

            We show the radio and X-ray luminosity of the sources with the 35 stars with {\it eROSITA} detection in the right panel of Fig. \ref{lxlbol}, with a colour bar showing the activity index. While the majority of the sample falls within the G\"{u}del–Benz fit, some of the stars fall below the fit, showing over-luminosities in the radio. In addition, this sample also shows a high activity index. Juxtaposing the two, we argue that active stars showing excess radio luminosity are highly chromospherically active. This shows that the same magnetic activity is responsible for the radio emission and chromospheric H\&K emission and that the gyrosynchrotron emission and activity index are correlated, which is no surprise since both mechanisms are non-thermal emission and are evidence of magnetic activity.

        \subsection{Time Series Investigation}
            \subsubsection{Period Analysis}

\begin{figure}
                \centering\includegraphics[width=\columnwidth]{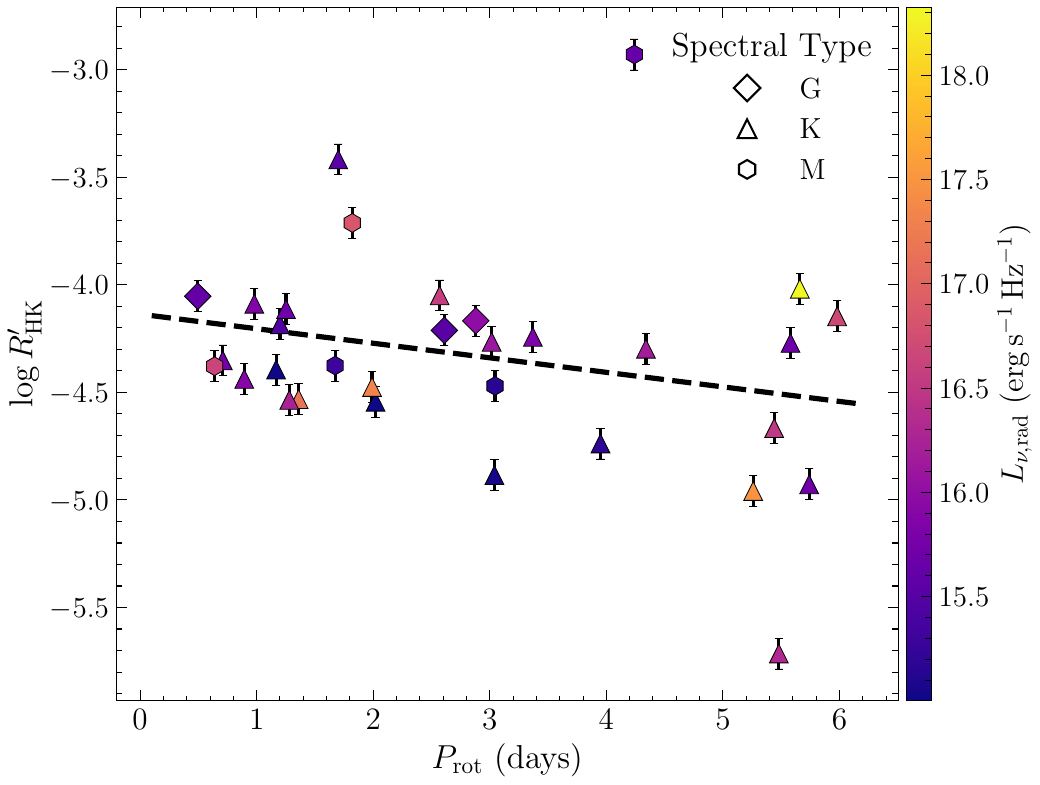}
                \caption{$R_{\mathrm{HK}}^{\prime}$ - $P_{\mathrm{rot}}$ relations for the 32 stars with orbital periods retrieved from literature and colour coded with the specific radio luminosity.}
                \label{porbhk}
            \end{figure}

            To complement the chromospheric and coronal diagnostics presented in the previous sections, we analysed the optical variability of the SMGPS chromospherically active candidates using publicly available time-domain photometric data. Optical variability provides a probe of the surface magnetic activity through rotational modulation by starspots and short-duration flare events, offering a photospheric counterpart to the radio and X-ray behaviour discussed earlier. We reviewed the literature and retrieved the rotation periods of the 39 stars observed with the SpUpNIC medium-resolution blue configuration. We found period information for 32 of these stars from \cite{Watson:2006, Samus':2017, Jayasinghe:2018, Kounkel:2022, Green:2023, Colman:2024, Lin:2024, Vach:2024}, many of which derived the periods using TESS light curves. Based on the retrieved periods of the 32 stars, we show the activity levels $R_{\mathrm{HK}}^{\prime}$ versus the rotation period $P_{\mathrm{rot}}$ in Fig. \ref{porbhk}. The stars span a broad range of rotation periods, $P_{\mathrm{rot}} \leq 6,\mathrm{d}$. No statistically significant correlation is evident across the sample. A linear fit yields a shallow negative slope of $-0.068 \pm 0.047$ with $R^{2}\approx 0.064$, indicating that the observed scatter is only weakly described by a linear trend. The weak negative trend is broadly consistent with the established rotation–activity relation, whereby chromospheric activity generally declines with increasing rotation period, although the substantial scatter indicates that rotation is not the sole factor governing the observed activity levels \citep{Noyes:1984, Wright:2011}. In most cases, activity levels decrease slightly as the rotation period increases. A linear fit indicates a weak negative correlation between $R_{\mathrm{HK}}^{\prime}$ and $P_{\mathrm{rot}}$.

            In addition, we retrieved the Asteroid Terrestrial-impact Last Alert System (ATLAS; \cite{Tonry:2018}), light curves from the Falling Star database to search for possible flaring cases for some of our candidates.

        \subsubsection{Flare Detection} 
            Stellar flares were identified using a multi-step procedure designed to separate short-duration impulsive events from longer-term stellar variability (e.g. rotational modulation due to starspots) and instrumental trends. For each light curve, the underlying quiescent stellar brightness was first estimated using Locally Weighted Scatterplot Smoothing (LOWESS; \citealt{Cleveland:1979}). This smoothing technique allows for the removal of slow variations caused by stellar rotation, evolving starspots, or instrumental systematics, while preserving rapid transient features such as flares \citep{Tonry:2018, Heinze:2018}.

            The flare signal was then isolated by computing residuals between the observed magnitudes and the LOWESS-derived baseline. To quantify the local photometric noise robust to outliers, we estimated the noise level within a sliding time window using the Median Absolute Deviation (MAD; \citealt{Rousseeuw:1993}), which was converted to an equivalent standard deviation under the assumption of Gaussian statistics. This local noise estimate was further smoothed using a Gaussian kernel to suppress small-scale fluctuations and ensure temporal coherence. Candidate flares were identified as residuals exceeding three times the locally estimated noise level.

            \begin{figure}
                \includegraphics[width=\columnwidth]{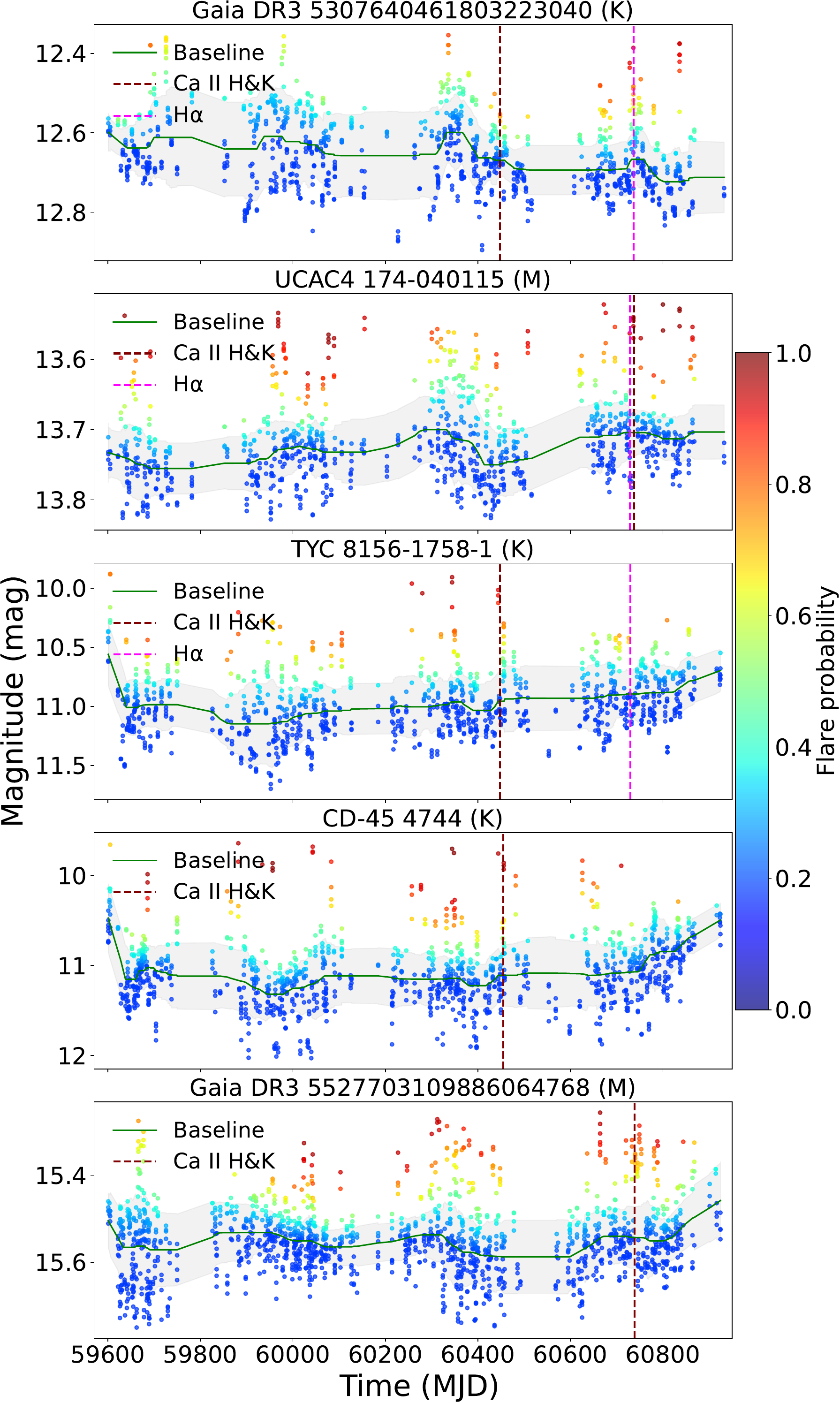}
                \caption{ATLAS light curves of  Gaia DR3 5307640461803223040, UCAC4 174-040115, TYC 8156-1758-1, CD-45 4744 and Gaia DR3 5527703109886064768. Each data point is colour-coded by flare probability. The baseline is the underlying trend of the light curve, whereas the shaded region represents the local noise envelope. The dashed vertical lines indicate the time of H$\alpha$ and Ca {\sc ii} H\&K detections.}
                \label{flare}
            \end{figure}

            We found that several stars in our sample exhibit clear flaring activity; five examples are shown in Fig. \ref{flare}  comprising Gaia DR3 5307640461803223040, UCAC4 174-040115, TYC 8156-1758-1, CD-45 4744 and Gaia DR3 5527703109886064768. UCAC4 174-040115 and Gaia DR3 5527703109886064768 are low-mass young stellar objects of spectral type M, whereas Gaia DR3 5307640461803223040, TYC 8156-1758-1 and CD-45 4744 are of K- spectral  types \citep{Zari:2018, GaiaCollaboration:2023}. For visual inspection, the original o-band light curves were annotated with the inferred baseline and local noise estimates, together with the identified flare events. Notably, some of the detected flares occur simultaneously with the detection of Ca {\sc ii} and H$\alpha$ lines. This temporal coincidence indicates that the photometric flares and the observed chromospheric line emission are manifestations of the same underlying magnetic activity processes. In particular, impulsive energy release during magnetic reconnection events heats the chromosphere and lower corona, producing enhanced Balmer and Ca {\sc ii} line emission, while simultaneously giving rise to broadband photometric flares \citep{Kowalski:2013, Mas-Buitrago:2025}. The alignment of flare epochs with chromospheric line detections, therefore, provides independent, multi-wavelength confirmation of enhanced stellar magnetic activity in these sources, and supports the interpretation that the observed emission lines are flare-driven rather than arising solely from steady-state chromospheric emission or accretion-related processes.

\section{Conclusion}
    Stellar magnetic activity in late-type stars results in phenomena that trigger thermal and non-thermal emission detected across multiple wavebands. Following the identification of radio and coronal X-ray stellar sources from the SMGPS and RASS/eRASS1, we have used the SAAO Lesedi 1-m and 1.9m telescopes, plus SALT, to carry out optical spectroscopic follow-up to search for chromospheric activity. In this work, we carried out spectroscopic observations of 99 candidate radio coronal stars, of which 46 show Ca {\sc ii} emission. Among these, 39 stars exhibit Ca {\sc ii} H\&K line emission identified from the SpUpNIC medium-resolution blue configuration. Of these, 23 are newly identified chromospherically active stars. Using SALT HRS observations, we found that 18 stars display Ca {\sc ii} infrared triplet (IRT) emission, 11 of which also show Ca {\sc ii} H\&K emission. A fraction of the stars also exhibit H$\alpha$ emission, detected in 14 of the 39 H\&K stars. The high-resolution SALT spectra reveal diverse H$\alpha$ morphologies, with several lines appearing broad and double-peaked, suggesting possible binarity. We compared the colour index,  $G_{BP} - G_{RP}$, and the activity index, $ R^{\prime}_{\mathrm{HK}}$ and observed no clear dependence on the colour index. However, the scatter is present among stars of similar colour, indicating that chromospheric activity is not determined solely by effective temperature or spectral type. Also, 37 out of 39 stars show activity levels above the Solar minimum. 
    
    The H$\alpha$ and Ca {\sc ii} H\&K fluxes exhibit positive log-linear correlations, with the strongest relationship observed in the simultaneous observations. In particular, the  tighter correlations obtained from simultaneous observations emphasise the importance of coordinated multi-wavelength observations when investigating stellar activity. The observed scatter in the chromospheric and multi-wavelength activity diagnostics further suggests that intrinsic stellar variability contributes significantly to the diversity of activity levels observed in this radio-selected sample. Future time-domain spectroscopic monitoring, ideally coordinated with radio observations, will help quantify the contributions of rotational modulation, activity cycles, and flaring, and further establish the connection between radio emission and chromospheric activity in magnetically active stars.

    From a multi-wavelength perspective, the radio and X-ray luminosities are within the ranges for late-type active stars, although in some cases their radio luminosities are in excess compared to the canonical G\"{u}del–Benz relationship. The excess could be attributed to flaring/variability, commonly observed in stellar radio transients, and can vary by up to one order of magnitude. The ratio of X-ray luminosity to bolometric luminosity, $L_X/L_{\rm bol}$, is predominantly within $-4 > L_X/L_{\rm bol}< -2.5$, a typical saturation ratio for coronally active stars. In comparing the rotation period information from TESS light curves with the activity levels derived from Ca {\sc ii} H\&K, we find no statistically significant correlation within the present sample, although the weak negative trend is qualitatively consistent with the canonical stellar rotation--activity relation.

    In summary, this work provides a spectroscopic characterisation of radio-selected late-type stars and demonstrates that their magnetic activity manifests across multiple wavelengths, from radio to optical to X-rays. The observed chromospheric activity, variability, and flaring behaviour reinforce the importance of coordinated and time-resolved observations for understanding the nature and evolution of magnetically active stellar populations.

\section*{Acknowledgements}
    The authors would like to thank Naomi Titus for her spectral data reduction codes.

    This work is based on research fully supported by the National Research Foundation's (NRF) South African Astronomical Observatory (SAAO) PhD Prize Scholarship. PJG and ODE are partly supported by the NRF SARChI grant 111692. DAHB acknowledges research support by the National Research Foundation.

     A part of this work is based on observations made with the Southern African Large Telescope (SALT) with the Large Science Programme on transients (2021-2-LSP-001 and 2024-2-LSP-001; PI: DAHB). Polish participation in SALT is funded by grant No. MEiN 2021/WK/01.

    The MeerKAT telescope is operated by the South African Radio Astronomy Observatory, which is a facility of the National Research Foundation, an agency of the Department of Science and Innovation.

    This work has made use of data from the European Space Agency (ESA) mission \href{https://www.cosmos.esa.int/Gaia}{\it Gaia}, processed by the {\it Gaia} Data Processing and Analysis Consortium \href{https://www.cosmos.esa.int/web/Gaia/dpac/consortium}{(DPAC)}. Funding for the DPAC has been provided by national institutions, in particular the institutions participating in the {\it Gaia} Multilateral Agreement. 

    This research has made use of the VizieR catalogue access tool, CDS, Strasbourg, France (\href{DOI:10.26093/cds/vizier}{DOI:10.26093/cds/vizier}). The original description of the VizieR service was published in 2000, A\&AS 143, 23.

    This work is based on data from eROSITA, the soft X-ray instrument aboard SRG, a joint Russian-German science mission supported by the Russian Space Agency (Roskosmos), in the interests of the Russian Academy of Sciences represented by its Space Research Institute (IKI), and the Deutsches Zentrum f\"ur Luft- und Raumfahrt (DLR). The SRG spacecraft was built by Lavochkin Association (NPOL) and its subcontractors, and is operated by NPOL with support from the Max Planck Institute for Extraterrestrial Physics (MPE). The development and construction of the eROSITA X-ray instrument was led by MPE, with contributions from the Dr. Karl Remeis Observatory Bamberg \& ECAP (FAU Erlangen-Nuernberg), the University of Hamburg Observatory, the Leibniz Institute for Astrophysics Potsdam (AIP), and the Institute for Astronomy and Astrophysics of the University of T\"ubingen, with the support of DLR and the Max Planck Society. The Argelander Institute for Astronomy of the University of Bonn and the Ludwig Maximilians Universit\"at Munich also participated in the science preparation for eROSITA. The eROSITA data shown here were processed using the eSASS software system developed by the German eROSITA consortium.

    This research made use of {\tt ASTROPY},\footnote{http://www.astropy.org} a community-developed core Python package for Astronomy \citep{AstropyCollaboration:2013, AstropyCollaboration:2018}. We also use {\tt NUMPY} \cite{Harris2020}, {\tt SCIPY} \citep{Virtanen2020}, {\tt PANDAS} \citep{reback2020pandas},  and {\tt MATPLOTLIB} \citep{Hunter:2007}. This research has made use of the SIMBAD database, operated at CDS, Strasbourg, France.

\section*{Data Availability}

    Information about the SARAO Galactic Plane Survey can be found in \cite{Goedhart:2024} and the compact source catalogue in \cite{Mutale:2025}. The coronal X-ray data from eROSITA eRASS1 are available in \cite{Freund:2024}. The Gaia DR3 data used in this work are available in the CDS Vizier database; the details can be found in \cite{ GaiaCollaboration:2023}. The SMGPS radio--X-ray star catalogue is available at the CDS (Centre de Donn\'ees astronomiques de Strasbourg) via \url{https://cdsarc.cds.unistra.fr/viz-bin/cat/J/A+A/707/A193}.



\bibliographystyle{mnras}
\bibliography{example} 




\appendix

\section{Ca {\sc ii} H\&K lines obtained with SpUPNIC medium-resolution blue configuration}

    \begin{figure*}
        \includegraphics[width=\textwidth]{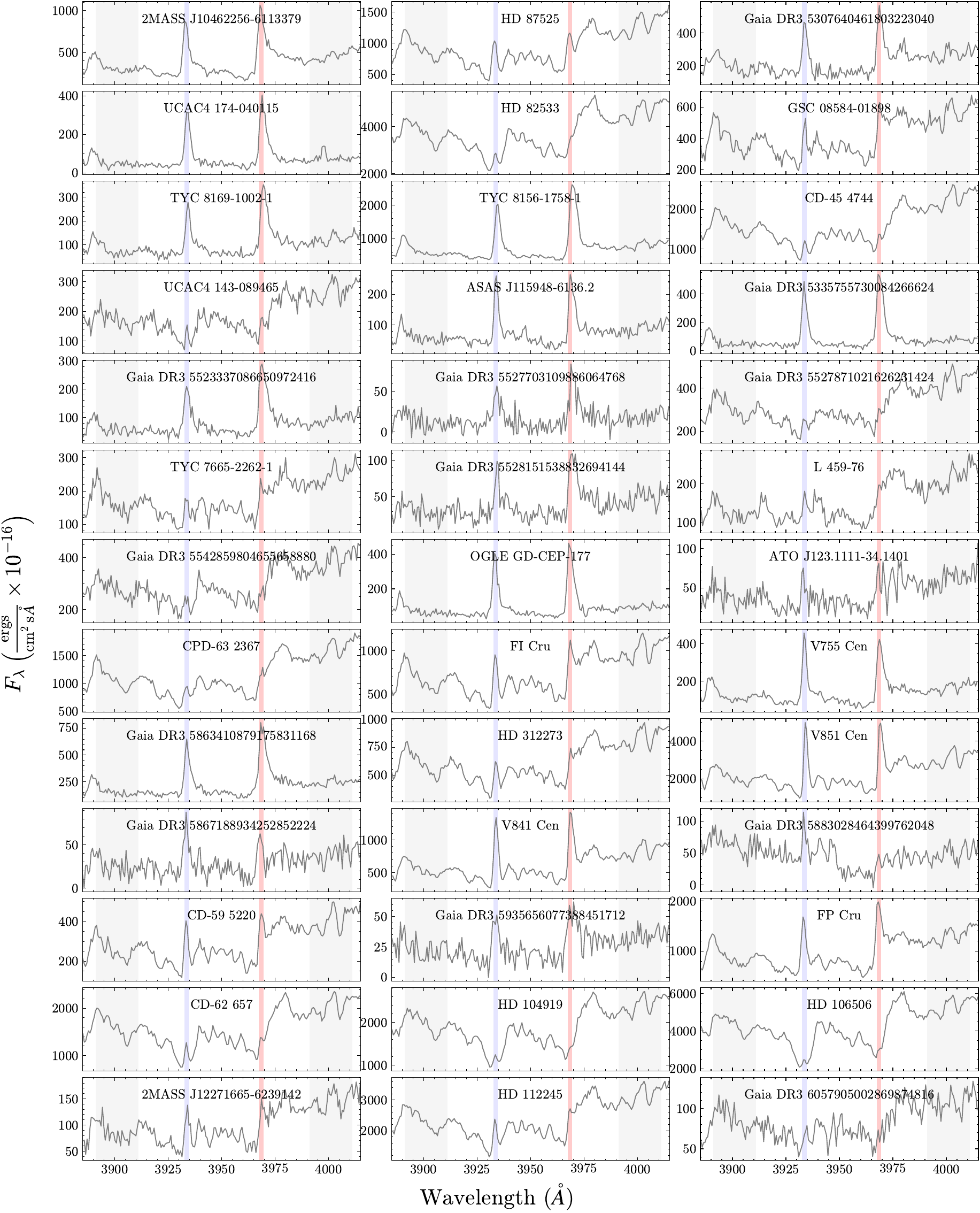}
        \caption{The spectra of the 39 SMGPS stars observed with SpUpNIC \sout{grating 4} {\bf medium-resolution blue configuration} showing the Ca {\sc ii} H \& K line region. Emission reversals in the line cores are visible across the sample, confirming their chromospheric activity. The shaded region in the spectra indicates the FWHM in the spectral line and the continuum region used in calculating the $S-$index.}
        \label{chmofig}
    \end{figure*}

    \begin{figure*}
        \includegraphics[width=0.6\textwidth]{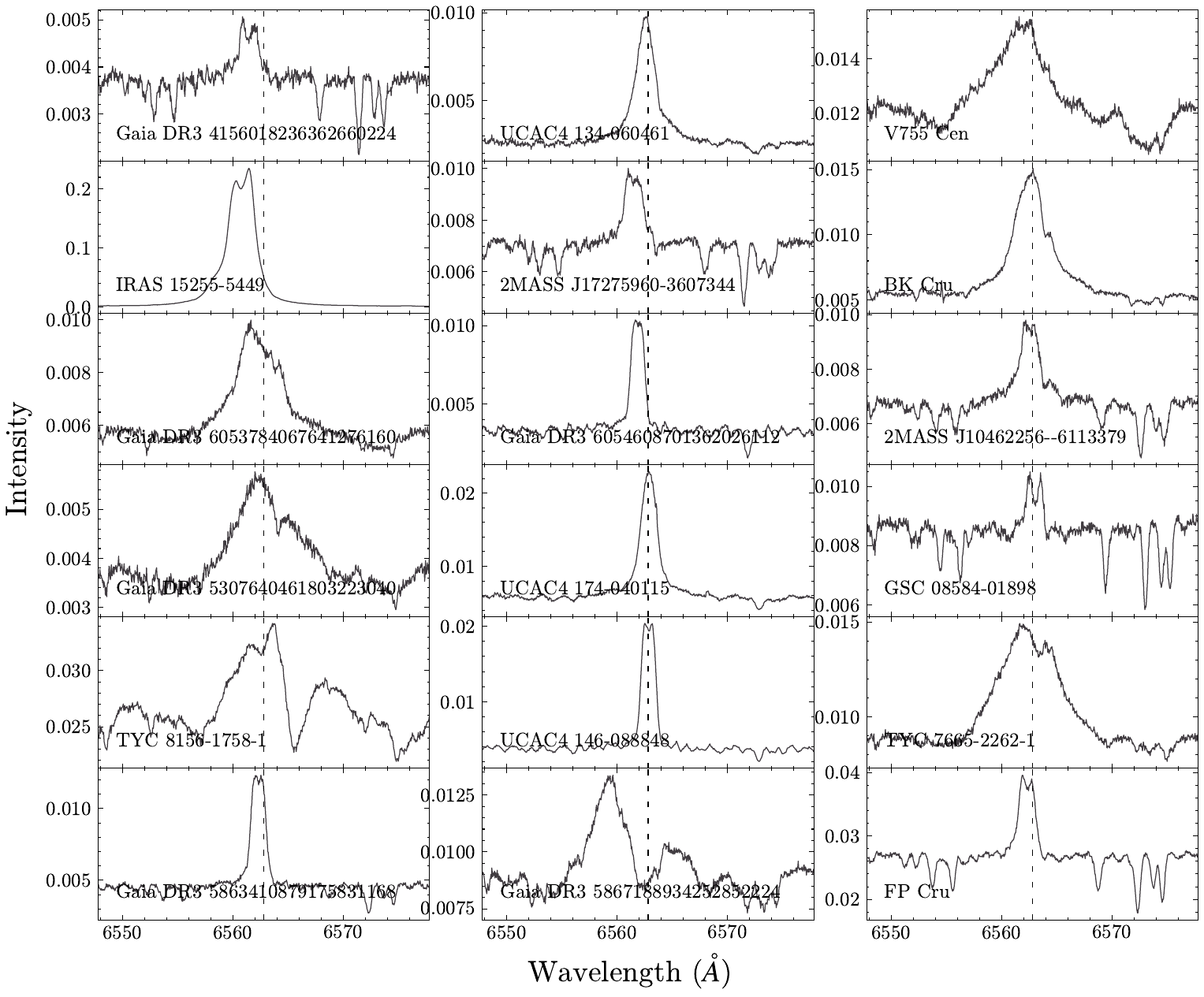}
        \caption{SALT HRS spectra showing H$\alpha$ line for 18 stars, a subset of our sample.}
        \label{hrsalpha}
    \end{figure*}

    \begin{figure*}
        \includegraphics[width=0.8\textwidth]{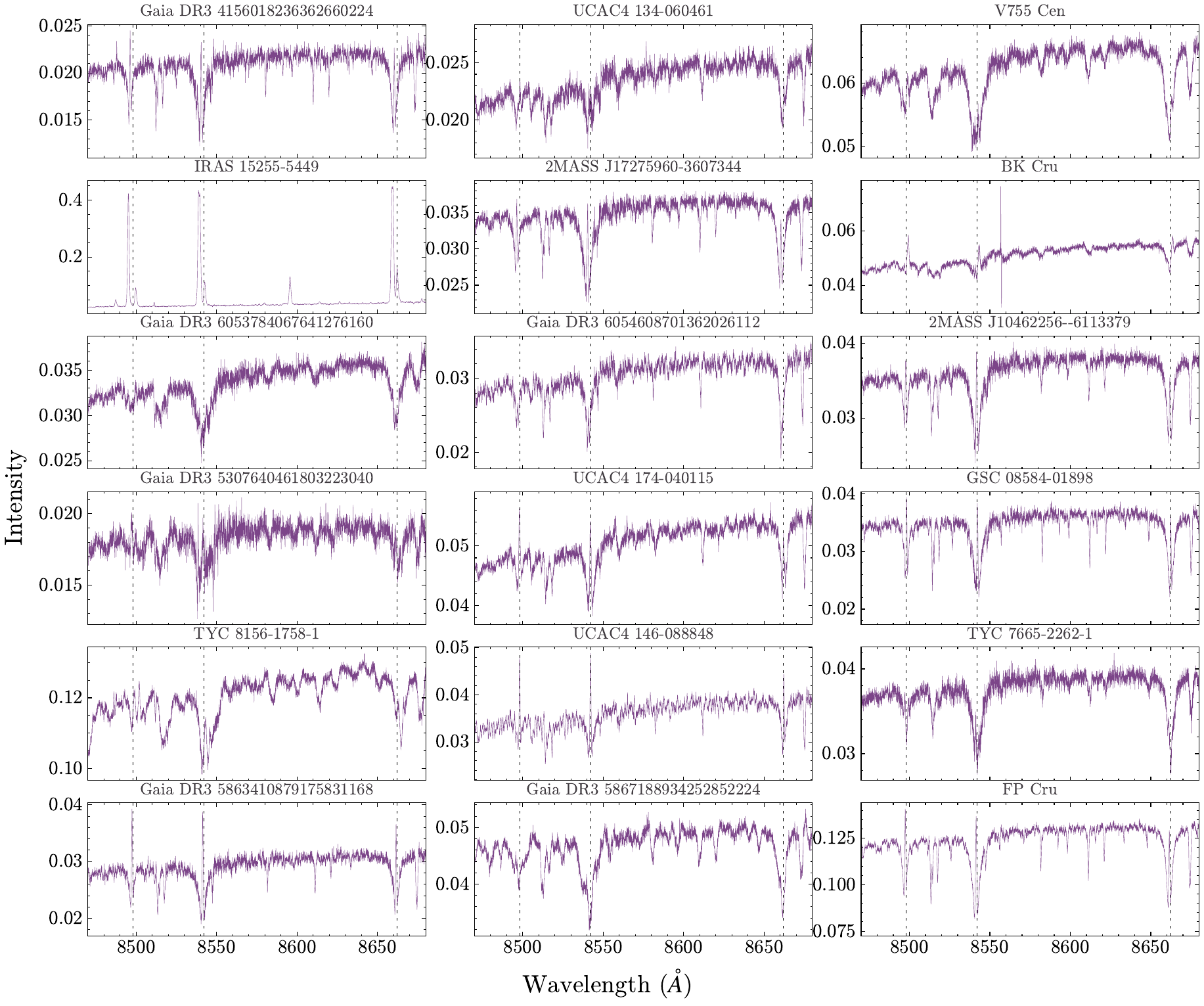}
        \caption{SALT HRS spectra showing Ca {\sc ii} IRT lines for 18 stars, a subset of our sample.}
        \label{hrscairt}
    \end{figure*}

\section{Log of Spectroscopic Observations conducted for this work}


    \begin{table*}
        \centering
        \caption{Log of \textit{Mookodi} Observations log}
        \begin{tabular}{lrlllr}
        \toprule
        
        Target Name & $G$ (mag) & Date & RA & Dec. & Exp. Time (s) \\
        
        \midrule
        TYC 8156-1758-1 & 10.10 & 2024-02-08 & 08:56:50.37 & --47:34:49.98 & 600.00 \\
        V755 Cen & 11.58 & 2023-08-19 & 13:05:37.78 & --62:12:12.31 & 300.00 \\
        Gaia DR3 5307640461803223040 & 12.95 & 2024-02-21 & 09:51:49.61 & --56:20:20.56 & 900.00 \\
        Gaia DR3 5523337086650972416 & 15.31 & 2023-05-26 & 08:37:00.62 & --43:21:33.68 & 720.00 \\
        \bottomrule
        
        \end{tabular}
        \label{mkdlog}
    \end{table*}


    \begin{table*}
        \centering
        \caption{Log of \textit{SpUPNIC} Observations with \sout{Grating 4} {\bf medium-resolution blue configuration.}}
        \begin{tabular}{lrlllr}
        \toprule
        Target Name & $G$ (mag) & Date & RA & Dec. & Exp. Time (s) \\
        
        \midrule
        OGLE GD-CEP-177 & 12.62 & 2025-03-03 & 08:08:38.041 & --36:04:00.274 & 1200.00 \\
        ATO J123.1111--34.1401 & 14.48 & 2025-03-03 & 08:12:26.684 & -34:08:24.591 & 2400.00 \\
        Gaia DR3 5542859804655658880 & 11.50 & 2024-05-19 & 08:21:24.508 & --36:29:07.828 & 900.00 \\
        TYC 7665-2262-1 & 12.04 & 2024-05-18 & 08:24:15.010 & -40:38:29.952 & 900.00 \\
        Gaia DR3 5527703109886064768 & 15.64 & 2025-03-05 & 08:25:02.512 & --40:54:41.569 & 3200.00 \\
        Gaia DR3 5528151538832694144 & 15.20 & 2025-03-05 & 08:25:46.225 & --39:11:16.069 & 3000.00 \\
        Gaia DR3 5527871021626231424 & 11.86 & 2024-05-25 & 08:28:13.162 & --39:54:36.219 & 900.00 \\
        L  459-76 & 11.13 & 2024-05-25 & 08:34:39.702 & -39:13:05.614 & 900.00 \\
        Gaia DR3 5523337086650972416 & 15.31 & 2025-03-05 & 08:37:00.615 & --43:21:33.683 & 3200.00 \\
        TYC 8156-1758-1 & 10.10 & 2024-05-18 & 08:56:50.366 & --47:34:49.977 & 600.00 \\
        CD-45  4744 & 8.65 & 2024-05-25 & 08:58:26.761 & -46:05:41.164 & 300.00 \\
        TYC 8169-1002-1 & 11.78 & 2024-05-19 & 09:10:07.890 & --48:37:42.975 & 900.00 \\
        HD  82533 & 8.09 & 2024-05-21 & 09:30:42.568 & -53:31:18.133 & 300.00 \\
        GSC 08584-01898 & 11.97 & 2024-05-18 & 09:31:25.422 & --53:14:36.611 & 900.00 \\
        Gaia DR3 5307640461803223040 & 12.95 & 2024-05-18 & 09:51:49.607 & --56:20:20.561 & 1200.00 \\
        UCAC4 174-040115 & 13.83 & 2025-03-04 & 09:52:18.721 & --55:16:37.493 & 3000.00 \\
        HD  87525 & 7.66 & 2024-05-17 & 10:03:52.563 & -56:11:22.884 & 240.00 \\
        2MASS J10462256--6113379 & 11.27 & 2025-03-04 & 10:46:22.545 & --61:13:37.750 & 1200.00 \\
        UCAC4 146-088848 & 13.40 & 2025-03-05 & 11:36:22.423 & --60:53:14.554 & 2800.00 \\
        UCAC4 143-089465 & 11.77 & 2024-05-21 & 11:43:33.712 & --61:28:28.093 & 900.00 \\
        ASAS J115948-6136.2 & 11.41 & 2024-05-17 & 11:59:49.869 & --61:36:24.623 & 1200.00 \\
        HD 104919 & 9.24 & 2024-05-25 & 12:04:48.863 & --64:09:55.427 & 300.00 \\
        HD 106506 & 8.28 & 2024-05-21 & 12:15:18.554 & --63:25:30.331 & 300.00 \\
        Gaia DR3 6057905002869874816 & 12.01 & 2024-05-25 & 12:19:36.892 & --61:52:56.021 & 600.00 \\
        FP Cru & 10.61 & 2024-05-17 & 12:21:30.783 & -64:03:52.688 & 720.00 \\
        2MASS J12271665--6239142 & 10.49 & 2024-05-17 & 12:27:16.641 & --62:39:14.177 & 720.00 \\
        CD-62   657 & 9.12 & 2024-05-25 & 12:28:25.395 & --63:20:58.738 & 300.00 \\
        CPD-63  2367 & 9.73 & 2024-05-22 & 12:36:38.978 & --63:44:43.520 & 720.00 \\
        FI Cru & 10.34 & 2024-05-18 & 12:44:34.806 & --63:31:46.232 & 720.00 \\
        HD 112245 & 9.16 & 2024-05-20 & 12:56:09.396 & --61:27:25.473 & 600.00 \\
        Gaia DR3 5863410879175831168 & 12.11 & 2025-03-05 & 12:58:26.705 & --62:30:54.023 & 1200.00 \\
        HD 312273 & 9.94 & 2024-05-25 & 13:02:47.023 & --62:13:59.053 & 360.00 \\
        V755 Cen & 11.58 & 2024-05-18 & 13:05:37.782 & --62:12:12.306 & 900.00 \\
        V851 Cen & 7.35 & 2024-05-18 & 13:44:00.921 & --61:21:59.173 & 240.00 \\
        Gaia DR3 5867188934252852224 & 12.70 & 2024-05-18 & 13:54:22.680 & --60:55:33.172 & 900.00 \\
        CD-59  5220 & 10.71 & 2024-05-18 & 14:23:47.426 & --59:35:26.055 & 720.00 \\
        V841 Cen & 8.38 & 2024-05-18 & 14:34:16.046 & --60:24:28.982 & 300.00 \\
        Gaia DR3 5883028464399762048 & 12.05 & 2024-05-22 & 15:29:26.318 & --57:17:01.942 & 900.00 \\
        Gaia DR3 5935656077388451712 & 12.57 & 2024-05-19 & 16:20:37.932 & --48:46:52.796 & 900.00 \\
        
        \bottomrule
        
        \end{tabular}
        \label{gr4log}

    \end{table*}



    \begin{table*}
        \centering
        \caption{Log of \textit{SpUPNIC} Observations with \sout{Grating 7} {\bf low-resolution wideband configuration.}}
        \begin{tabular}{lrlllr}
        \toprule
        Target Name & $G$ (mag) & Date & RA & Dec. & Exp. Time (s) \\
        
        \midrule
        TYC 7665-2262-1 & 12.04 & 2024-05-27 & 08:24:15.09 & --40:38:31.23 & 420.00 \\
        TYC 8169-1002-1 & 11.78 & 2024-05-27 & 09:10:07.87 & --48:37:42.90 & 420.00 \\
        2MASS J10462256--6113379 & 11.27 & 2024-05-28 & 10:46:22.17 & --61:13:37.63 & 360.00 \\
        UCAC4 146-088848 & 13.40 & 2024-05-27 & 11:36:22.42 & --60:53:14.23 & 540.00 \\
        ASAS J115948--6136.2 & 11.41 & 2024-05-27 & 11:59:49.76 & --61:36:25.64 & 420.00 \\
        FP Cru & 10.61 & 2024-05-27 & 12:21:30.77 & --64:03:53.74 & 360.00 \\
        FI Cru & 10.34 & 2024-05-28 & 12:44:34.55 & --63:31:46.56 & 360.00 \\
        V851 Cen & 7.35 & 2024-05-28 & 13:44:00.77 & --61:21:59.55 & 60.00 \\
        V841 Cen & 8.38 & 2024-05-29 & 14:34:15.77 & --60:24:29.38 & 90.00 \\
        CD-59  5220 & 10.71 & 2024-05-29 & 14:23:47.27 & --59:35:26.21 & 360.00 \\
        \bottomrule
        
        \end{tabular}
        \label{gr7log}
    \end{table*}





    \begin{table*}
    \centering
        \caption{Log of \textit{SALT HRS} Observation Log}
        \begin{tabular}{lrlllr}
        \toprule
        Target Name & $G$ (mag) & Date & RA & Dec. & Exp. Time (s) \\
        \midrule
        
            TYC 7665-2262-1 & 12.04 & 2025-03-05 & 08:24:15.010 & --40:38:29.952 & 500.00 \\
            TYC 8156-1758-1 & 10.10 & 2025-02-24 & 08:56:50.366 & -47:34:49.977 & 500.00 \\
            GSC 08584-01898 & 11.97 & 2025-02-24 & 09:31:25.422 & --53:14:36.611 & 500.00 \\
            Gaia DR3 5307640461803223040 & 12.95 & 2025-03-03 & 09:51:49.607 & --56:20:20.561 & 1500.00 \\
            UCAC4 174-040115 & 13.83 & 2025-02-23 & 09:52:18.721 & --55:16:37.493 & 2000.00 \\
            2MASS J10462256--6113379 & 11.27 & 2025-02-23 & 10:46:22.545 & --61:13:37.750 & 500.00 \\
            UCAC4 146-088848 & 13.40 & 2025-02-26 & 11:36:22.423 & --60:53:14.554 & 1500.00 \\
            UCAC4 134-060461 & 13.14 & 2025-02-28 & 11:47:29.196 & --63:12:54.995 & 1500.00 \\
            FP Cru & 10.61 & 2025-02-28 & 12:21:30.783 & --64:03:52.688 & 500.00 \\
            BK Cru & 13.08 & 2025-03-10 & 12:21:53.191 & --63:47:33.533 & 1500.00 \\
            Gaia DR3 6053784067641276160 & 12.11 & 2025-02-23 & 12:23:15.932 & --63:16:50.121 & 750.00 \\
            Gaia DR3 6054608701362026112 & 12.59 & 2025-02-23 & 12:23:58.595 & --62:43:20.020 & 1000.00 \\
            Gaia DR3 5863410879175831168 & 12.11 & 2025-03-03 & 12:58:26.705 & --62:30:54.023 & 750.00 \\
            V755 Cen & 11.58 & 2025-03-09 & 13:05:37.782 & --62:12:12.306 & 500.00 \\
            Gaia DR3 5867188934252852224 & 12.70 & 2025-03-15 & 13:54:22.680 & --60:55:33.172 & 1275.00 \\
            IRAS 15255-5449 & 13.73 & 2025-03-17 & 15:29:24.930 & --54:59:21.588 & 2000.00 \\
            2MASS J17275960--3607344 & 11.57 & 2025-03-03 & 17:27:59.598 & --36:07:34.531 & 500.00 \\
            Gaia DR3 4156018236362660224 & 13.53 & 2025-03-18 & 18:26:30.361 & -09:45:38.016 & 1500.00 \\
        \bottomrule
        
        \end{tabular}
        \label{hrslog}
    \end{table*}


\bsp	
\label{lastpage}
\end{document}